\documentclass[sigconf]{acmart}

\AtBeginDocument{%
  }

\copyrightyear{2026}
\acmYear{2026}
\setcopyright{cc}
\setcctype{by}
\acmConference[CCS '26]{Proceedings of the 2026 ACM SIGSAC Conference on Computer and Communications Security}{November 15--19, 2026}{The Hague, Netherlands}
\acmBooktitle{Proceedings of the 2026 ACM SIGSAC Conference on Computer and Communications Security (CCS '26), November 15--19, 2026, The Hague, Netherlands}
\acmDOI{10.1145/3830454.3846620}
\acmISBN{979-8-4007-2871-6/2026/11}

\usepackage{cprotect}
\usepackage{lipsum}
\usepackage{tcolorbox}
\usepackage{xspace}
\usepackage{enumitem}
\usepackage{graphicx}
\usepackage{multirow}
\usepackage{listings}
\usepackage{soul}
\usepackage[dvipsnames,table,xcdraw]{xcolor}
\usepackage{array}
\newcolumntype{C}[1]{>{\centering\arraybackslash}p{#1}}

\usepackage{filecontents}
\usepackage{tikz}
\usetikzlibrary{arrows.meta, positioning, calc}
\usepackage{tikzpagenodes}
\usepackage{subcaption}
\usepackage{booktabs}
\usepackage{xurl}
\usepackage{appendix}
\usepackage[absolute,overlay]{textpos}
\usepackage{adjustbox}
\usepackage{makecell}

\usepackage[noabbrev,capitalize]{cleveref}
\acmBadgeR[https://www.acm.org/publications/policies/artifact-review-and-badging-current]{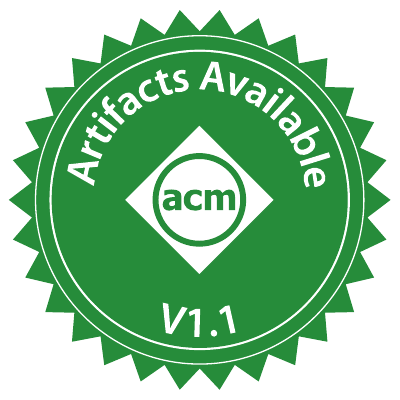}
\acmBadgeR[https://www.acm.org/publications/policies/artifact-review-and-badging-current]{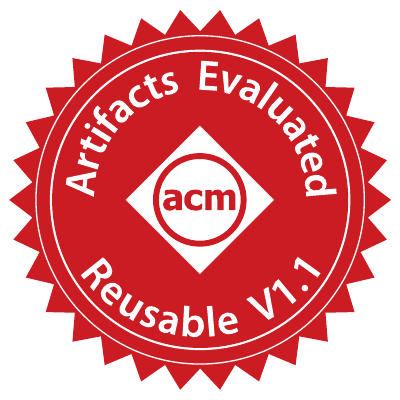}
\acmBadgeR[https://www.acm.org/publications/policies/artifact-review-and-badging-current]{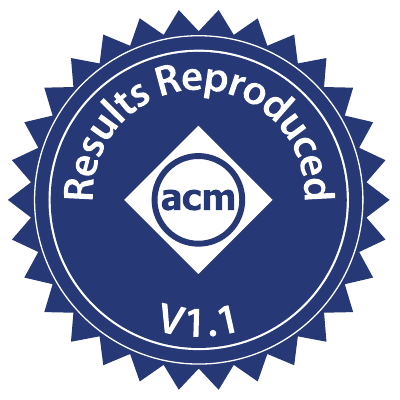}
\begin{document}

\title{\papername: Amplifying Rowhammer Attacks via Non-Uniform Patterns to Exploit ECC-Protected GPUs}



\author{Chris S. Lin}
\affiliation{
\institution{University of Toronto}
\city{Toronto}
\country{Canada}}
\email{shaopenglin@cs.toronto.edu}

\author{Joyce Qu}
\affiliation{
\institution{University of Toronto}
\city{Toronto}
\country{Canada}}
\email{joyce.qu@mail.utoronto.ca}

\author{Aditya Rajeev}
\affiliation{
\institution{University of Toronto}
\city{Toronto}
\country{Canada}}
\email{arajeev@cs.toronto.edu}

\author{Gururaj Saileshwar}
\affiliation{
\institution{University of Toronto}
\city{Toronto}
\country{Canada}}
\email{gururaj@cs.toronto.edu}




\begin{abstract}
GDDR memory in GPUs is vulnerable to Rowhammer attacks, where rapid memory accesses induce bit flips in adjacent cells, enabling data tampering and privilege escalation. However, prior GPU Rowhammer attacks trigger only tens to hundreds of bit flips, orders of magnitude fewer than CPU attacks, severely limiting their practical impact. 
This gap stems from the reliance of existing GPU Rowhammer attacks on uniform hammering patterns that activate aggressor and decoy rows equally, which results in low hammering intensity for aggressor rows.

We present \papername{}, a high-intensity Rowhammer attack on NVIDIA GPUs leveraging non-uniform hammering. \papername{} reverse engineers GPU memory-access coalescing behavior to enable non-uniform hammering patterns on GPUs, that activate aggressor rows more intensely than decoy rows. Additionally, by identifying refresh instances when in-DRAM mitigations are applied, it constructs longer attack patterns that escape mitigation across refresh intervals, further increasing hammering intensity.
Together, these techniques yield 500$\times$ to 23,500$\times$ more bit flips than prior GPU Rowhammer attacks, across several NVIDIA GPUs (A4000, A4500, A5000, A6000), reaching bit flip rates close to state-of-the-art CPU Rowhammer attacks.
\papername{} also enables the first Rowhammer exploits on ECC-protected GPUs, inducing uncorrectable double and triple bit flips, making denial-of-service and privilege-escalation attacks practical even on GPUs with ECC enabled.
\end{abstract}


\begin{abstract}
GDDR memory in GPUs is vulnerable to Rowhammer attacks, where rapid memory accesses induce bit flips in adjacent cells, enabling data tampering and privilege escalation. However, prior GPU Rowhammer attacks trigger only tens to hundreds of bit flips, orders of magnitude fewer than CPU attacks, severely limiting their practical impact. 
This gap stems from the reliance of existing GPU Rowhammer attacks on uniform hammering patterns that activate aggressor and decoy rows equally, which results in low hammering intensity for aggressor rows.

We present GPUThor, a high-intensity Rowhammer attack on NVIDIA GPUs leveraging non-uniform hammering. GPUThor reverse engineers GPU memory-access coalescing behavior to enable non-uniform hammering patterns on GPUs, that activate aggressor rows more intensely than decoy rows. Additionally, by identifying refresh instances when in-DRAM mitigations are applied, it constructs longer attack patterns that escape mitigation across refresh intervals, further increasing hammering intensity.
Together, these techniques yield 500X to 23,500X more bit flips than prior GPU Rowhammer attacks, across several NVIDIA GPUs (A4000, A4500, A5000, A6000), reaching bit flip rates close to state-of-the-art CPU Rowhammer attacks.
GPUThor also enables the first Rowhammer exploits on ECC-protected GPUs, inducing uncorrectable double and triple bit flips, making denial-of-service and privilege-escalation attacks practical even on GPUs with ECC enabled.
\end{abstract}





\newcommand{\TODO}[1]{{\footnotesize\color{red}[TODO: #1]}}
\newcommand{\red}[1]{{\color{red}#1}}
\newcommand{\blue}[1]{{\color{blue}#1}}

\newcounter{observationcounter}
\renewcommand{\theobservationcounter}{\arabic{observationcounter}} 

\newtcolorbox{observation}[1][]{
    colback=black!7, colframe=black!85!black, coltitle=black,
    left=1mm, right=1mm, top=1mm, bottom=1mm, boxrule=0.5mm,
    #1
}

\newcommand{\papername}{GPUThor}

\newcommand{\FlipProbDef}{\text{flip probability}\xspace}
\newcommand{\TRC}{\text{tRC}\xspace}
\newcommand{\REF}{\text{REF}\xspace}
\newcommand{\TRH}{$\text{T}_{\text{RH}}$\xspace}
\newcommand{\HCfirst}{$\text{HC}_{\text{first}}$\xspace}
\newcommand{\ACT}{\text{ACT}\xspace}
\newcommand{\TREFI}{\text{tREFI}\xspace}
\newcommand{\TREFW}{\text{tREFW}\xspace}
\newcommand{\TRFC}{\text{tRFC}\xspace}
\newcommand{\TRP}{\text{tRP}\xspace}

\newcommand{\observationref}[1]{Obs.~\ref{#1}}


\maketitle

\section{Introduction}

\label{sec:intro}
Rowhammer is a read-disturbance vulnerability in DRAM that enables attackers to induce bit flips in memory cells by rapidly activating neighboring rows~\cite{Rowhammer2014}. Such attacks have been shown to enable data tampering, sandbox escapes, and privilege escalation exploits~\cite{ProjectZeroRowhammer, TRRespass, Blacksmith, SMASH, ZenHammer,Eccploit}. While these were first discovered over a decade ago in CPU-based DDR memories, in the past year, several new Rowhammer attacks have been demonstrated on GPU-based GDDR memories~\cite{gpuhammer,gpubreach,gddrhammer,geforge}. GPUHammer~\cite{gpuhammer} first discovered Rowhammer bit flips on NVIDIA A6000 GPUs with GDDR6 memory, 
and used them to degrade ML model accuracy. Subsequent works~\cite{gpubreach,gddrhammer, geforge} further showed that Rowhammer on GPUs can even lead to system-wide privilege escalation, establishing GPU Rowhammer attacks as a potent threat to system security.

Despite these recent advances, the number of bit flips observed by recent GPU Rowhammer attacks remains {\it orders of magnitude lower} than what CPU-based attacks routinely achieve~\cite{heckel2026flippyram,Blacksmith}, limiting their practicality. 
As shown in \cref{tab:comparison}, GPUHammer~\cite{gpuhammer} reported just 2 bit flips per DRAM bank on an A6000 GPU (16 flips per GB), while GeForge~\cite{geforge} and GPUBreach~\cite{gpubreach} reported 2.2 and 5.6 bit flips per bank (18 and 45 flips per GB) respectively on the same GPU.
GDDRHammer~\cite{gddrhammer} improved this to 94.8 bit flips per bank (758 flips per GB) using double-sided hammering patterns. In contrast, CPU-based Rowhammer attacks like Blacksmith~\cite{Blacksmith} have demonstrated up to 550,000 flips per GB on DDR4 memories, orders of magnitude higher than the best GPU Rowhammer attacks.
Due to the low bit flip rates on GPUs, enabling ECC on GDDR-based GPUs, which provides SECDED-level protections and corrects single-bit errors, effectively mitigates prior Rowhammer attacks.
Accordingly, NVIDIA recommends enabling ECC as the primary defense against GPU Rowhammer attacks~\cite{NVIDIARowhammerNotice}.
In this paper, we bridge the gap between GPU and CPU attacks, and demonstrate GPU Rowhammer attacks triggering up to \textit{23,500$\times$ more bit flips} than prior attacks, and show that \textit{ECC is an insecure mitigation for GPUs}.

\vspace{0.1in}
\noindent
\textbf{Limitations of Prior Work.} The fundamental limitation of prior GPU Rowhammer attacks is that they perform \emph{uniform} hammering: all rows in the attack pattern, both aggressor rows (adjacent to the target victim) and decoy rows (used to evade in-DRAM mitigations), are activated at equal rates. Thus, a significant fraction of each refresh interval is spent hammering decoy rows rather than aggressors. In contrast, CPU Rowhammer attacks like Blacksmith~\cite{Blacksmith} employ \emph{non-uniform} hammering, activating aggressors at much higher intensity relative to decoys while still evading mitigations, dramatically increasing bit flips. No prior GPU Rowhammer attack has successfully utilized non-uniform hammering, leaving the threat of Rowhammer on GPUs severely underestimated.\footnote{While 
GeForge~\cite{geforge} \textit{claims} to perform non-uniform hammering, it has a bit-flip rate (2.2 per bank) similar to GPUHammer (2 per bank)~\cite{gpuhammer} and lower than GPUBreach (5.6 per bank)~\cite{gpubreach} which both use uniform hammering; it is likely that its access patterns collapse to uniform hammering due to memory request coalescing (\textbf{C1}).}
However, 
adapting non-uniform hammering to GPUs is non-trivial and requires addressing a few key challenges (\textbf{C1-C3}).

\smallskip
\noindent\textbf{C1. Memory Access Coalescing.} Unlike CPUs, we discover that GPUs have optimizations that aggressively coalesce memory accesses at multiple levels within warps and at the memory controller to amortize memory bandwidth. This coalescing prevents the fine-grained control over activations that non-uniform hammering requires: repeated accesses intended to hammer aggressors at higher intensity may be merged into a single DRAM activation, collapsing a naive non-uniform pattern back to a uniform one.

\noindent\textbf{C2. Unknown Mitigation Instances.} Non-uniform hammering can be more effective with the knowledge of \emph{when} in-DRAM mitigations sample aggressors and issue mitigative refreshes, so that true aggressors can be hammered intensely outside the sampling window.
In CPU DRAMs, these intervals are known to be aligned to multiples of \TREFI~\cite{Blacksmith, SMASH}. However, the sampling and mitigation instances in GDDR6 memories have not been characterized, leaving attackers unable to design effective non-uniform patterns.

\noindent\textbf{C3. Defeating ECC.} Even if hammering intensity is amplified, workstation-class GPUs (e.g., A6000) support SECDED ECC that can correct single-bit errors and detect double-bit errors in GDDR6 memory. 
This makes it difficult for Rowhammer attacks to be successful in the presence of ECC.

\smallskip
\noindent \textbf{Our Approach.}
We introduce \emph{\papername{}}, the first high-intensity, non-uniform Rowhammer attack on NVIDIA GPUs.
Our approach is enabled by reverse-engineering the micro-architectural behavior of GPU memory accesses and GDDR6 in-DRAM mitigations.

First, to address memory request coalescing (C1), we systematically characterize the behavior of repeated memory requests on GPUs, within and across warps. We find that while repeated requests within a single warp are aggressively coalesced at the memory controller, requests across warps are typically not coalesced. Moreover, when targeting different cachelines from across warps, requests further have minimal interference at the cache level. Leveraging this, we design hammering kernels that distribute repeated accesses to a row across warps on \emph{independent cachelines} within a row, ensuring distinct repeated activations. This yields a 2--3$\times$ increase in hammering intensity over prior works~\cite{gpuhammer,gpubreach,gddrhammer,geforge}.

Second, to overcome the unknown mitigation instances (C2), we reverse-engineer the Target Row Refresh (TRR) behavior on Ampere GPUs with GDDR6 memories. We discover that mitigations are issued approximately once every {72 \TREFI{}s}, rather than once per \TREFI{} as assumed by prior works~\cite{gpuhammer,gpubreach}. Leveraging this, we develop non-uniform attack patterns spanning up to 6 \TREFI{}s, 
with the first five \TREFI{}s interleaving repeated aggressor accesses with decoy accesses; the final \TREFI{} contains only decoy accesses.
This increases hammering intensity, i.e.,  activation rates per aggressor, to nearly 6.6$\times$ that of prior uniform patterns~\cite{gpuhammer}.

\smallskip
\noindent \textbf{Hammering Campaigns.}
We evaluate \papername{} on four Ampere-class GPUs (A4000, A4500, A5000, A6000), a broader set than prior works~\cite{gpuhammer,gpubreach,gddrhammer,geforge}. Across four banks per GPU with ECC disabled, we observe \textbf{72,000–377,000 bit flips per GB}, up to 500$\times$ higher than GDDRHammer~\cite{gddrhammer} and 23,500$\times$ higher than GPUHammer~\cite{gpuhammer}.
These flip rates approach the DDR4 bit flip rates with CPU attacks like Blacksmith (550,000 flips/GB)~\cite{Blacksmith}, indicating comparable vulnerability.
At these bit-flip rates, even SECDED ECC, assumed to be applied at 16-byte granularity in workstation GPUs~\cite{IMTSullivan}, is insufficient. Across these GPUs (four banks each), at 16-byte granularity, we discover \textbf{387 double-bit flips} that SECDED ECC can detect but cannot correct (DUE), and \textbf{2 triple-bit flips} that the ECC can neither detect nor correct, resulting in silent data corruption (SDC); the most vulnerable GPU, the A5000, accounts for 306/387  double-bit flips and 2/2 triple-bit flips.
We thus surpass the challenge of defeating ECC (C3) that limits prior GPU Rowhammer attacks.

\begin{table}[t]

\centering
\caption{Number of bit flips across GPU Rowhammer attacks. \papername{} induces up to 500$\times$ and 23{,}500$\times$ more flips per GB than prior works, GDDRHammer and GPUHammer.}
\label{tab:comparison}
\small
\begin{adjustbox}{max width=\linewidth}
\begin{tabular}{llrrr}
\toprule
\textbf{Attack} & \textbf{GPU} & \textbf{Flips/Bank} & \textbf{Flips/GB} & \textbf{Ratio} \\
\midrule
GPUHammer*~\cite{gpuhammer}       & RTX A6000 & 2.0    & 16      & 1$\times$ \\
GeForge*~\cite{geforge}           & RTX A6000 & 2.2    & 18    & 1.1$\times$ \\
GPUBreach*~\cite{gpubreach}       & RTX A6000 & 5.6    & 45    & 2.8$\times$ \\
GDDRHammer*~\cite{gddrhammer}     & RTX A6000 & 94.8   & 758   & 47$\times$ \\
\midrule
\multirow{4}{*}{\textbf{\shortstack[l]{\papername{}\\(our work)}}}
    & RTX A6000 & \textbf{14{,}311} & \textbf{114{,}488} & \textbf{7{,}155$\times$} \\
l    & RTX A5000 & \textbf{23{,}597} & \textbf{377{,}552} & \textbf{23{,}597$\times$} \\
    & RTX A4500 & \textbf{4{,}689}  & \textbf{75{,}024}  & \textbf{4{,}689$\times$} \\
    & RTX A4000 & \textbf{4{,}548}  & \textbf{72{,}768}  & \textbf{4{,}548$\times$} \\
\bottomrule
\end{tabular}
\end{adjustbox}
\smallskip
*Bit flip counts for prior works are from their respective papers.
\end{table}

\smallskip
\noindent \textbf{Implications for ECC.}
We launch campaigns with ECC enabled on the A6000 GPU (our A4000-5000 GPUs are cloud-based, where the provider does not give the option to enable ECC). 
On the A6000 with ECC enabled, we trigger detectable uncorrectable errors (DUE), i.e., double-bit errors, causing ECC failures and GPU crashes, at an average rate of 1 DUE per hour; on the more vulnerable A5000 GPU, we estimate it would suffer one DUE per 20 minutes on average. Each DUE renders the GPU unusable and requires a GPU reset or a system reboot, enabling denial-of-service attacks, as recovery on cloud systems can take up to 20 minutes per DUE~\cite{cui2025storygpuscharacterizingresilience}.

Crucially, we discover that enabling ECC does not prevent privilege escalation exploits; we observe exploitable multi-bit errors with ECC enabled. Triple-bit errors occurring while ECC was enabled cause the SECDED code to mis-correct, leading to silent data corruption (SDC) that allows the corrupted data to be consumed \emph{without triggering a DUE that kills the GPU}. Moreover, we discover that even double-bit DUEs are exploitable, since DUEs are serviced lazily in NVIDIA GPUs, leaving a $\sim$10\,ms time window between DUE detection and the GPU being killed, during which the corrupted data is consumed by the attacker's GPU kernel. Exploiting these properties of SDC and DUEs, we demonstrate that \emph{privilege escalation attacks} are indeed feasible on an A6000 GPU \emph{even with ECC enabled}. These results show that NVIDIA-recommended defenses such as ECC are easily bypassed and stronger defenses are needed to mitigate GPU Rowhammer attacks.

\smallskip
\noindent \textbf{Contributions.}
This paper makes the following contributions:
\begin{enumerate}
    \item We demonstrate \papername{}, the first non-uniform, high-inte\-nsity Rowhammer attack on NVIDIA GPUs, inducing 500$\times$ to 23{,}500$\times$ more bit flips than prior works.
    \item We reverse-engineer the coalescing behavior of GPU memory accesses and TRR mitigation instances to enable effective non-uniform, multi-\TREFI{} hammering patterns on GPUs.
    \item Using \papername{} on GPUs like A4000, A4500, A5000, and A6000, we discover 72{,}000 to 377{,}000 bit flips/GB in GDDR6 DRAM with ECC disabled, and hundreds of multi-bit flips (387 double and 2 triple bit flips) uncorrectable by ECC.
    \item We demonstrate the first Rowhammer-based denial-of-servi\-ce and privilege escalation attacks on ECC-protected GPUs.
\end{enumerate}

\paragraph{Responsible Disclosure.}
We disclosed our findings to NVIDIA on 29 April 2026, and also to the major cloud vendors (Google, Microsoft, AWS). NVIDIA requested an embargo until 25th August 2026, when they planned to release a security notice.


\section{Background}
\label{sec:background}

In this section, we discuss the threat model, the memory system, and the GDDR6 DRAM on NVIDIA GPUs, GPU Rowhammer attacks, and error correction codes (ECC) as a defense on NVIDIA GPUs.

\subsection{Threat Model}
We assume an unprivileged attacker who can launch CUDA kernels on an NVIDIA GPU with GDDR6 memory, such as A4000--A6000, popularly used in workstations and the cloud. 
The attacker can co-locate with another victim user on a time-shared GPU in the cloud~\cite{GKE} and try to tamper with its data using Rowhammer, or run from an unprivileged process in a single-tenant setting and attempt privilege escalation to root by tampering GPU Page Tables via Rowhammer~\cite{gpubreach, geforge, gddrhammer}. Like prior GPU Rowhammer attacks~\cite{gpuhammer, gpubreach, geforge, gddrhammer}, the attacker can reverse-engineer the GPU virtual-address-to-DRAM bank and row mappings using timing side-channels and hammer neighbors of victim rows. 
We assume the IOMMU is enabled, similar to prior work~\cite{gpubreach}.

\subsection{GPU Memory System}
\label{subsec:gddr6_arch}

\begin{figure}[t]
\centering
\includegraphics[width=\linewidth,height=\paperheight,keepaspectratio]{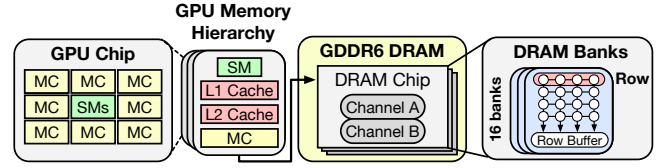}
\caption{Memory system in NVIDIA GPUs with GDDR6 DRAM. SMs issue memory requests that are serviced by the L1 or L2 cache or the memory controllers (MC). Each memory controller is connected to a GDDR6 DRAM with two channels, and each DRAM chip has 16 banks; within a bank, DRAM cells are organized as rows, all sharing a common row buffer.}
\label{fig:dram_org}
\end{figure}

\noindent\textbf{GPU Memory Hierarchy.} As shown in \cref{fig:dram_org}, NVIDIA GPUs have an array of Streaming Multiprocessors (SMs) that issue memory requests through on-chip L1/L2 caches to a set of on-die memory controllers (MC). Threads within a warp (a group of 32 threads) issue memory instructions in lockstep, and each MC in workstation GPUs (A4000--6000) is connected to independent GDDR6 DRAM channels~\cite{JEDEC-GDDR6}.  
The GPU memory system can \emph{coalesce} multiple requests to the same L1 cacheline within a warp to amortize its cost~\cite{nvidia_cuda_programming}, although coalescing behavior at other levels (L2 cache and memory controller) is undocumented. As we show later (\S\ref{sec:coalescing}), such coalescing has implications for mounting Rowhammer attacks.

\smallskip
\noindent\textbf{GDDR6 Organization.} As shown in \cref{fig:dram_org}, a GDDR6 device is organized as a hierarchy of channels, chips, banks, and rows~\cite{JEDEC-GDDR6}.  Each channel is split into two independent subchannels (Channel~A and Channel~B), and each chip contains 16 banks; each bank contains thousands of rows of cells storing charge (1s and 0s), and all rows in a bank share a common \emph{row buffer}. Data is accessed at the granularity of a row: an \ACT{} command opens the addressed row and latches it into the row buffer, column reads/writes operate on the buffered row, and a \texttt{PRE} command precharges the bitlines before another row in the same bank can be activated.  

\smallskip
\noindent\textbf{Refresh.} DRAM cells leak charge and must be periodically refreshed to preserve their contents. Thus, the memory controller issues a \REF{} command once every refresh interval \TREFI{} ($\le$1.9\,$\mu$s in GDDR6), which 
refreshes a subset of rows, so that every row is refreshed within the refresh window \TREFW{} (32\,ms in GDDR6)~\cite{JEDEC-GDDR6}. 
\REF{} commands also provide the DRAM the time to issue Rowhammer mitigation refreshes~\cite{TRRespass,UncoverRowhammer};
so \TREFI{} is also the typical time unit for constructing Rowhammer access patterns~\cite{Blacksmith, SMASH,gpuhammer}.


\subsection{CPU Rowhammer Attacks and Defenses}
\label{subsec:gpu_rowhammer}

\noindent\textbf{Rowhammer.} Rowhammer is a read-disturbance vulnerability in DRAM in which repeatedly activating (``hammering'') a row causes charge leakage in electrically adjacent rows, ultimately flipping bits in those victim cells~\cite{Rowhammer2014}.  The minimum number of activations to a row to induce a bit flip is called the \emph{Rowhammer threshold} (\HCfirst{}).  Rowhammer access patterns can target a \emph{victim} row by hammering one (\emph{single-sided}) or both (\emph{double-sided}) of its immediate neighbors, also called \textit{aggressors}. On CPUs, Rowhammer has been used for a variety of exploits, including privilege escalation, sandbox escape, and cryptographic key recovery~\cite{ProjectZeroRowhammer, Drammer, RowhammerJS, SMASH, ZenHammer, Blacksmith, Eccploit}.

\smallskip
\noindent\textbf{Target Row Refresh (TRR).} Modern DRAM chips, starting from DDR4, deploy in-DRAM mitigations against Rowhammer, collectively referred to as Target Row Refresh (TRR). TRR samples frequently activated rows and issues a \emph{mitigative refresh} to their neighbors to reverse the charge leakage in victim rows~\cite{TRRespass, UncoverRowhammer}. However, as TRR trackers have finite capacity, subsequent attacks~\cite{TRRespass,Blacksmith} evaded TRR by hammering \emph{aggressor} rows interleaved with a larger set of \emph{decoy} rows, overwhelming the tracker so the aggressors escape mitigation. While TRRespass~\cite{TRRespass} used \textit{uniform} patterns that hammered aggressors and decoys at the same rate, Blacksmith~\cite{Blacksmith} developed \textit{non-uniform} patterns that hammer aggressors at a higher rate than decoys, increasing hammering intensity and flipping bits on a wider range of devices, while also significantly increasing the number of bit flips (up to 550K flips per GB in DDR4 devices~\cite{Blacksmith}). 
As TRR mitigations align with multiples of \TREFI{}~\cite{UncoverRowhammer}, such attacks synchronize the hammering patterns with \REF{} commands~\cite{SMASH}.
Recent attacks~\cite{phoenix,ZenHammer} also demonstrate that such non-uniform attack patterns induce bit flips in DDR5 devices with on-die ECC.

\begin{figure*}[t]
\centering
\includegraphics[width=\textwidth]{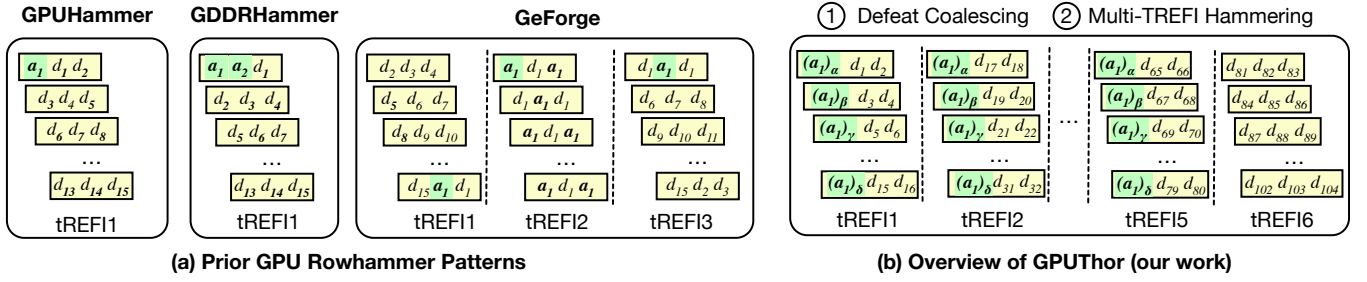}
\caption{(a) Prior GPU Rowhammer patterns ($a$ - aggressor row, $d$ - decoy row, aggressor activations in green):
GPUHammer and GDDRHammer hammer aggressors uniformly (1-2$\times$  activations/\TREFI{}) with single \TREFI{} patterns; GeForge suffers coalescing with 3 \TREFI{} patterns.
(b) Overview of \papername{}. It performs non-uniform hammering using 6 \TREFI{} patterns with 6.6$\times$ activations per aggressor per \TREFI{}; it avoids coalescing by repeating aggressors across warps and unique cachelines ($\alpha$, $\beta$, $\gamma$, $\delta$).} 
\label{fig:RH_patterns}
\end{figure*}

\subsection{GPU Rowhammer Attacks}
\textbf{GPUHammer}~\cite{gpuhammer} first showed that GPU Rowhammer attacks are practical on an NVIDIA A6000 GPU with GDDR6 memory. By reverse-engineering the mappings of virtual addresses to GDDR6 banks and rows, it developed multi-thread and multi-warp hammering kernels for hammering GDDR6 memories.
As shown in \cref{fig:RH_patterns}~(a), by using $n$-sided patterns, i.e., one aggressor and $n-1$ decoys per \TREFI{}, and \textit{uniform} hammering (1$\times$ aggressor activation per \TREFI{}, same as decoys) like TRRespass~\cite{TRRespass}, it defeated TRR to induce bit flips in GDDR6 memories.
GPUHammer used these bit flips to tamper with ML-model weights and degrade the model accuracy. 
\textbf{GPUBreach}~\cite{gpubreach} used similar uniform hammering patterns to induce bit flips, and tampered with GPU page table entries to enable system-wide privilege escalation.

\textbf{GDDRHammer}~\cite{gddrhammer} and \textbf{GeForge}~\cite{geforge} also demonstrated privilege escalation attacks, with IOMMU disabled.
GDDRHammer extends GPUHammer's uniform hammering from single-sided to double-sided, as shown in \cref{fig:RH_patterns}~(a), providing 34$\times$ more bit flips~\cite{gddrhammer}.
While it evaluated non-uniform patterns accessing aggressors with higher intensity and multi-\TREFI{} patterns, it found the most bit flips with its \textit{uniform} pattern (single \TREFI{}, 2$\times$ aggressor activations per \TREFI{}).
GeForge \textit{claims} to use a \textit{non-uniform} hammering pattern that spans three \TREFI{} periods, accessing an aggressor row up to 13 times across 3 \TREFI{}s. 
However, as it repeats accesses to the same row within a warp, based on our analysis (\S\ref{sec:coalescing}), these accesses get \emph{coalesced} to a few activations per \TREFI{}, resulting in approximately uniform hammering like GPUHammer.
The number of bit flips with GeForge is just 1.1$\times$ that of GPUHammer~\cite{gpuhammer} and less than GPUBreach~\cite{gpubreach}, which use uniform hammering.

As summarized in \cref{tab:comparison}, all four prior GPU Rowhammer attacks use largely \emph{uniform} hammering patterns and only have tens of bit flips per bank, much lower than prior CPU attacks~\cite{Blacksmith, heckel2026flippyram}. Given the low bit-flip rates in GPUs so far, enabling ECC on GPUs has been sufficient to protect against these attacks~\cite{gpuhammer,gpubreach,gddrhammer,geforge}.

\subsection{Error Correction Codes on GPUs}
\label{subsec:gpu_ecc}

 \noindent\textbf{ECC.} Workstation-class NVIDIA GPUs (e.g., A4000--A6000) with GDDR6 memories support error correction codes (ECC) that provide SECDED (single-error-correct, double-error-detect) level protection~\cite{IMTSullivan}. When ECC is enabled, prior analysis~\cite{IMTSullivan,gpuhammer} suggests that these GPUs store a 2\,B ECC for every 32\,B data (a 6.25\% memory overhead). Thus, at the granularity of 16\,B data, a 1\,B SECDED code can \emph{correct} any single bit flip (single-bit error, SBE) and \emph{detect} but not correct any two-bit flip (double-bit error, DBE). A DBE results in a detectable, uncorrectable error (DUE), while with $\ge 3$ flips, the data can get mis-corrected, causing silent data corruption (SDC).
SECDED ECC in CPU DRAM has been shown to be insecure against Rowhammer attacks, with exploits shown on ECC-DIMMs in DDR3~\cite{Eccploit} and DDR4~\cite{eccfail}, and on DDR5 with on-die ECC~\cite{phoenix,ZenHammer}.
However, thus far, on GDDR6, enabling ECC mitigates all existing GPU Rowhammer attacks.
NVIDIA also recommends enabling ECC on GPUs to defend against GPU Rowhammer attacks~\cite{NVIDIARowhammerNotice}.


\smallskip
\noindent\textbf{Error Management.} 
In the event of a correctable SBE, the NVIDIA GPU silently repairs the SBE and the kernel continues executing; in the background, a counter tracking corrected errors is incremented, which is readable by an unprivileged user process via \texttt{nvidia-smi}.
When a DUE is detected on Ampere GPUs, all the running GPU kernels are aborted and the GPU is rendered unusable until it is reset~\cite{nvidia_error_management}.
Additionally, a DUE triggers \emph{row remapping}, in which the affected row is permanently retired and replaced with a spare row on the next GPU reset; if the pool of spare rows is exhausted, the device raises a \textit{row-remapping failure} flag, indicating the device qualifies for RMA (return-merchandise-authorization), marking it as defective~\cite{nvidia_error_management}.  
This occurs in the event of 9 DUEs in a bank or after 512 DUEs across the entire GPU memory~\cite{nvidia_error_management}.

\section{Overview of \papername{}}
\label{sec:overview}

The goal of \papername{} is to engineer Rowhammer attacks on NVIDIA GPUs with high hammering intensity, capable of producing orders of magnitude more bit flips than prior GPU attacks.
We seek to evaluate the real vulnerability of these GPUs and whether ECC-based defenses recommended by NVIDIA~\cite{NVIDIARowhammerNotice} are truly secure. 
We achieve this by enabling \emph{non-uniform} hammering on GPUs inspired by state-of-the-art CPU Rowhammer attacks~\cite{Blacksmith, phoenix, ZenHammer}.

\smallskip
\noindent\textbf{Approach.}
To mount high-intensity GPU Rowhammer attacks via non-uniform patterns, \papername{} relies on two key building blocks:


\smallskip
\noindent\textit{(1) Defeating GPU memory request coalescing (\S\ref{sec:coalescing}).}
Since the GPU memory subsystem aggressively coalesces accesses within warps, achieving repeated DRAM activations for the same row is challenging for GPU Rowhammer attacks, unlike on CPUs. We characterize coalescing of memory requests at three levels (within a warp, across warps to the same address, and across warps to different cachelines of the same row) and find that the third regime reliably preserves repeated activations. Therefore, our hammering kernels distribute repeated accesses to an aggressor row across multiple warps {\it and} across unique cachelines of that row, enabling repeated DRAM activations and non-uniform hammering.


\smallskip
\noindent\textit{(2) Multi-\TREFI{} non-uniform hammering (\S\ref{sec:multitrefi}).}
While single \TREFI{} hammering enables non-uniform patterns, there is a limit on how much the intensity of aggressors can be increased with a single \TREFI{} pattern, as the placement of decoy rows with respect to TRR sampling instances becomes constrained. To address this, we develop multi-\TREFI{} non-uniform hammering patterns. 
We characterize the mitigation instances on GDDR6 chips in Ampere GPUs, constructing non-uniform patterns spanning multiple \TREFI{}s and using the resulting bit flip reproducibility as a side channel to learn TRR mitigation frequency.  
We find that TRR mitigation is performed approximately once every 72 \TREFI{}s, rather than once per \TREFI{}~\cite{gpuhammer,gpubreach}.
We enable non-uniform hammering patterns with length up to 6 \TREFI{}s and 6.6 aggressor activations per \TREFI{}.

Combining (1) and (2) gives the \papername{} pattern shown in \cref{fig:RH_patterns}~(b): a hammering pattern spanning 6 \TREFI{}s, with the first 5 \TREFI{}s having a total of 40 aggressor activations, and the final \TREFI{} reserved for decoys to overwhelm TRR. 
Next, we describe these building blocks for our hammering pattern  (\S\ref{sec:coalescing}, \S\ref{sec:multitrefi}).

\section{Defeating GPU Memory Request Coalescing}
\label{sec:coalescing}
In CPU Rowhammer attacks, hammering intensity can be easily increased by adding repeated memory accesses within a \TREFI{}, inducing more activations. However, GPUs can perform memory request coalescing, making this challenging. While NVIDIA documentation~\cite{nvidia_cuda_programming} suggests that coalescing of requests occurs within warps at the L1 cache level, there is insufficient detail on request coalescing at the L2 cache or activation coalescing at the memory controller, making triggering repeated \ACT{}s difficult. Hence, we first reverse engineer the memory request coalescing on NVIDIA GPUs and then defeat it to craft non-uniform Rowhammer patterns.

\subsection{Reverse Engineering Approach}
For our experiments, we reverse engineer the virtual address to DRAM bank and row mappings, similar to prior work, GPUHammer~\cite{gpuhammer}.
For all of our accesses, we use an \texttt{ld.volatile} followed by \texttt{discard} to ensure accesses skip the L1 and L2 caches respectively. 
We use an A5000 GPU for our experiments, but we validate that the memory access coalescing behavior is similar across GPUs, including A4000-A6000 (Ampere, GDDR6), L4 (Ada, GDDR6), and A30 (Ampere, HBM2). 

\begin{figure}[t]
\centering

\includegraphics[width=3.3in,height=\paperheight,keepaspectratio]{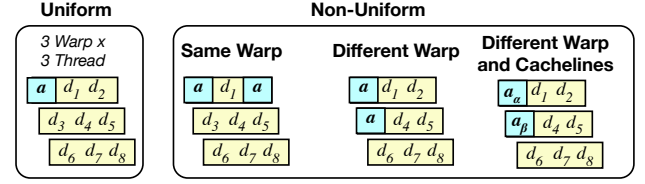}
\caption{Non-uniform patterns (repeated accesses to aggressor rows) used to study request coalescing behavior on GPUs.}
\label{fig:coalescing_patterns}
\end{figure}

As our baseline, we use a uniform 9-sided hammering pattern, similar to prior work GPUHammer~\cite{gpuhammer}, consisting of 3 Warps $\times$ 3 Threads, that accesses 9 unique rows: 1 aggressor ($a$) and 8 decoy rows ($d_1$-$d_{8}$). We then generate non-uniform patterns with 2$\times$ and 3$\times$ intensity for row $a$, by replacing one or two decoy rows in the pattern with repeated aggressor row ($a$) accesses, as shown in \cref{fig:coalescing_patterns}.
To validate whether the non-uniform patterns produce $2\times$ or $3\times$ activations for the repeated aggressor row ($a$), we measure the \textit{time per round} for each pattern, i.e., the time to complete one round of 9 accesses in the pattern, and check if the time is consistent with 9  \ACT{}s, similar to the baseline uniform pattern. We measure the time per round for both the uniform patterns and the non-uniform patterns, as shown in \cref{fig:coalesced_timing}.

To reverse engineer the coalescing behavior within and across warps, we perform the additional accesses to the row $a$ either within the same warp as the original access, or all in different warps. We also use addresses from different cachelines, at least 128\,B away  ($a_\alpha$, $a_\beta$, $a_\gamma$) to repeat access to row $a$, in same-warp and different-warp configurations, to study hammering of unique cachelines in a row.


\subsection{Coalescing \textit{Within} a Warp}

\begin{figure}[ht]
\centering
\includegraphics[width=3.3in,height=\paperheight,keepaspectratio]{"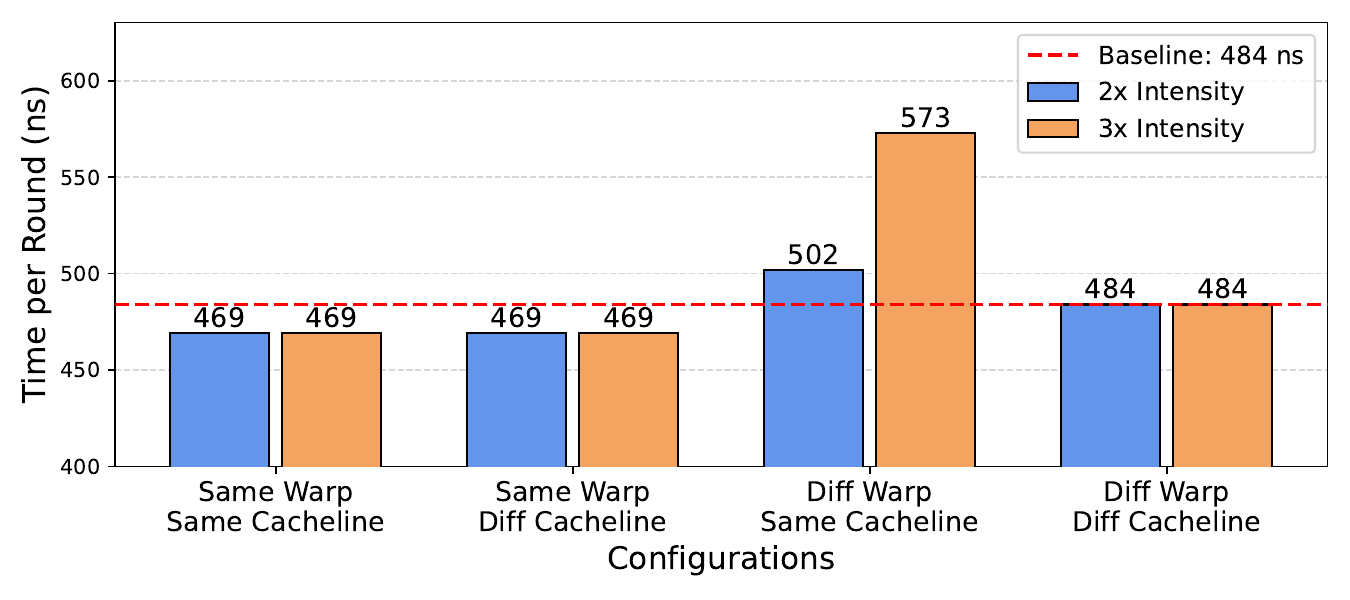"}
\caption{Average time for a round of hammering with $2-3\times$ repeated accesses to an aggressor row, accessed within or across warps, using addresses from the same or different cachelines in the row. The baseline is the time for a uniform pattern hammering 9 rows without repetition.}
\label{fig:coalesced_timing}
\end{figure}

\cref{fig:coalesced_timing} shows the time per round for the non-uniform patterns hammering additional aggressor rows, with intensity $2\times$ and $3\times$ within the same warp, compared to the baseline uniform hammering (1$\times$ intensity).
While the uniform pattern has a time per round of 484\,ns (for 9 \ACT{}s), the non-uniform patterns with additional accesses in the same warp have a lower time per round of 469\,ns, indicating that the pattern did not generate 9 ACTs. 
This is regardless of whether the additional accesses occur to the same cacheline or different cachelines in the same row. 
This suggests that the additional requests do not induce unique \ACT{}s to row $a$, and they likely get coalesced at the memory controller.

\begin{observation}\refstepcounter{observationcounter}
\textbf{Observation~\theobservationcounter.} Repeated requests to a row \textbf{within a warp} get coalesced at the memory controller, and \textbf{do not generate unique \ACT{}s}, and are unsuitable for non-uniform hammering.
\end{observation}

\subsection{Repeated Requests \textit{Across} Warps}
\label{sec:coalescing_across_warps}
In \cref{fig:coalesced_timing}, in non-uniform patterns where the aggressor row is repeatedly hammered from different warps, we see a timing spike when the repeated access is from the same cacheline, jumping to 502\,ns and 573\,ns per round for $2\times$ and $3\times$ intensity hammering, from the baseline 484\,ns.
This is because \texttt{discard} from the first warp, which evicts the cacheline from the L2 cache, interferes with the repeated request for the cacheline from the second warp, which causes the latency of these accesses to increase far beyond the \ACT{} latency.
Thus, while repeated accesses across warps may not get coalesced, the timing spike due to the \texttt{discard}'s contention across warps can cause the synchronization of the pattern with \TREFI{} to be disturbed, making it unsuitable for hammering.

In contrast, when repeated accesses from different warps are to different cachelines of the row, 128\,B apart, the timings are similar to those of the uniform hammering (484\,ns), ensuring that there is no coalescing of requests across warps.  
This indicates that these non-uniform patterns likely have the same number of \ACT{}s in total as the uniform hammering pattern, and additional \ACT{}s for the aggressor row, $a$. Across different warp/thread configurations, the time per round is closest to the baseline when the repeated accesses across warps are to different cachelines, even when it does not exactly match the baseline, indicating the ACTs are preserved.
\begin{observation}\refstepcounter{observationcounter}
\textbf{Observation~\theobservationcounter.} Repeated requests from \textbf{different warps} for \textbf{different cachelines} in a row are not coalesced and generate \textbf{unique \ACT{}s}, suitable for non-uniform hammering.
\end{observation}

\subsection{Non-Uniform Hammering in a Single \TREFI{}}
Using the non-uniform pattern from \cref{sec:coalescing_across_warps}, we perform hammering campaigns. 
As a baseline, we use a 24-sided uniform pattern (8 warps $\times$ 3 threads) that has 1 aggressor and 23 decoy rows accessed uniformly within a \TREFI{}, like GPUHammer~\cite{gpuhammer}: this overwhelms the 16-entry TRR sampler~\cite{gpuhammer}.
We construct non-uniform single-\TREFI{} patterns with $2\times$, $3\times$, and $4\times$ intensity by repeating aggressor accesses across different warps using distinct cachelines.

\cref{tab:rephammer} reports bit flips with these non-uniform patterns on an A5000 GPU across four random DRAM banks. 
The $1\times$ uniform baseline (GPUHammer) induces, on average, only 21 bit flips, while the $2\times$ non-uniform pattern yields an average of 32.5 bit flips, a $1.5\times$ increase. 
However, higher intensity patterns ($3\times$ and $4\times$) show a decrease in flips compared to the $2\times$ patterns.
This suggests that while non-uniform hammering with higher intensity can improve bit flips, scaling beyond $2\times$ intensity within a single \TREFI{} is challenging.
As aggressor intensity increases, the positions of decoys become constrained within the \TREFI{} window, reducing their chance of being sampled by TRR and their likelihood of fooling TRR. 
To alleviate this, we explore non-uniform patterns spanning multiple \TREFI{}s.

\begin{table}[hbt]
\centering
\caption{Number of bit flips on an A5000 GPU using non-uniform, single-\TREFI{} patterns with $2\times$, $3\times$ and $4\times$ intensity, compared to the baseline $1\times$ intensity uniform pattern~\cite{gpuhammer}.}
\label{tab:rephammer}
\small
\begin{adjustbox}{max width=\linewidth}
\begin{tabular}{lcC{1.4cm}C{1.4cm}C{1.4cm}}
\toprule
& \textbf{Baseline~\cite{gpuhammer}} &\multicolumn{3}{c}{\textbf{Non-Uniform, Single-\TREFI{}}}\\
\cmidrule(lr){2-2} \cmidrule(lr){3-5}
& \textbf{$1\times$} & \textbf{$2\times$} & \textbf{$3\times$} & \textbf{$4\times$} \\
\midrule
Bank 1 & 22 & 29 & 20 & 12 \\
Bank 2 & 15 & 25 & 16 & 11 \\
Bank 3 & 27 & 49 & 35 & 27 \\
Bank 4 & 20 & 27 & 23 & 15 \\
\midrule
\textbf{Average} & 21 & 32.5 (1.5$\times$) & 23.5 (1.1$\times$)  & 16.3 (0.8$\times$)\\

\bottomrule
\end{tabular}
\end{adjustbox}
\end{table}







\begin{observation}\refstepcounter{observationcounter}
\textbf{Observation~\theobservationcounter.} Non-uniform single-\TREFI{} hammering patterns increase the number of bit flips by $1.5\times$ by increasing intensity. But successful single-\TREFI{} patterns are limited to $2\times$ intensity, prompting the exploration of multi-\TREFI{} patterns.
\end{observation}

\section{Multi-tREFI Hammering on GPUs}
\label{sec:multitrefi}

To overcome the limitation of single-\TREFI{} patterns, we investigate non-uniform hammering across multiple \TREFI{}s. Our goal is to increase aggressor intensity while continuing to evade TRR. For this, using a known bit flip in \cref{tab:rephammer}, we try to reproduce it using multi-\TREFI{} patterns while systematically exploring the design space of non-uniform patterns.
First, keeping aggressor activation counts fixed, we evaluate whether spreading them out over multiple \TREFI{}s retains the bit flip (\S\ref{sec:multref_inc_trefi}). 
Next, we replace decoys with aggressor activations to study the potential to increase intensity (\S\ref{sec:multref_inc_intensity}).


\subsection{Increasing Pattern Lengths to \texorpdfstring{$n$}{n} \TREFI{}s} \label{sec:multref_inc_trefi}



\begin{figure}[ht]
\centering
\includegraphics[width=3in,height=\paperheight,keepaspectratio]{"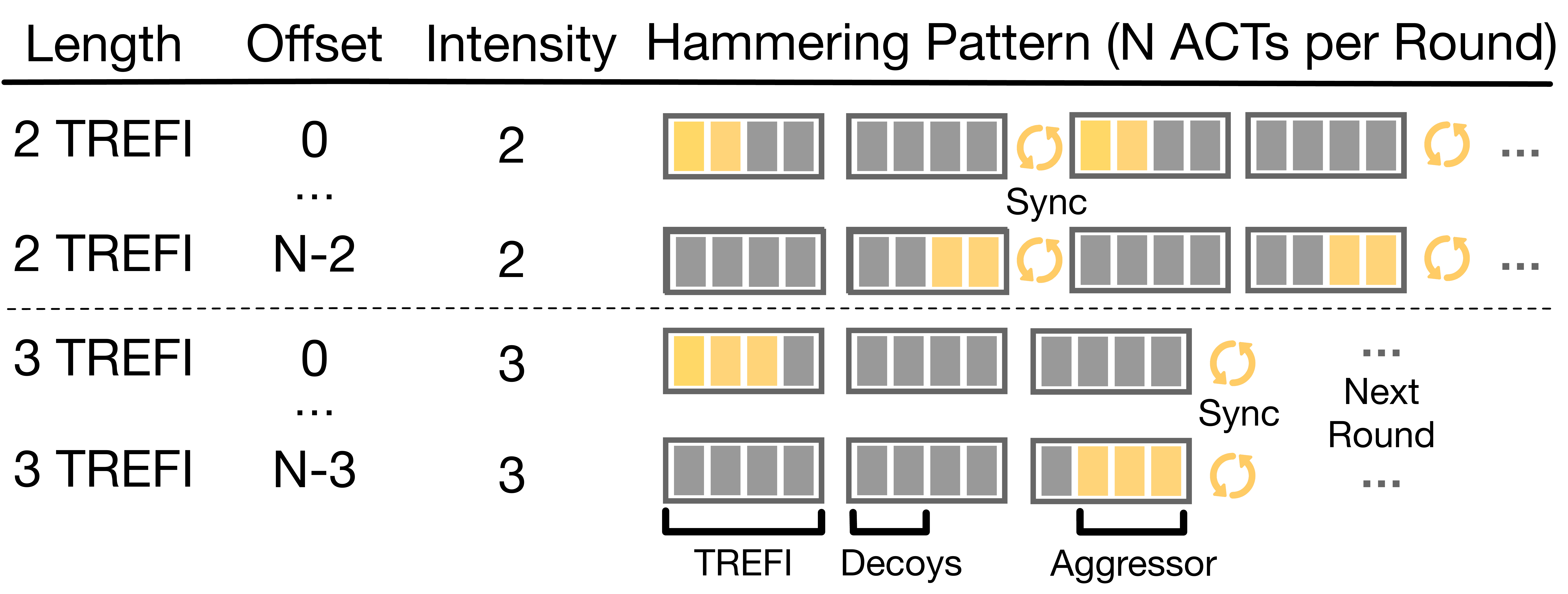"}
\caption{Hammering non-uniform patterns as the length ($n$-\TREFI{}s) increases, keeping intensity per \TREFI{} fixed.}
\label{fig:rephammer_vary_trefi}
\end{figure}

We reverse-engineer TRR mitigation frequency using non-uniform multi-\TREFI{} patterns by increasing pattern lengths from 1 to $n$ \TREFI{}s and measuring the \textit{\FlipProbDef{}} for the pattern over 50 trials, 
defined as the percentage of trials in which the hammering pattern is able to successfully reproduce a given bit flip.
As shown in \cref{fig:rephammer_vary_trefi}, our $n$-\TREFI{} hammering pattern consists of an aggressor receiving $n$ \ACT{}s in a round (aggressor intensity of $n$) and the remaining \ACT{}s target randomly chosen decoy rows (one \ACT{} per decoy).
We keep aggressor intensity per \TREFI{} constant across pattern lengths, ensuring the total aggressor activations remain the same. 
After one round of $n$-\TREFI{}s, the pattern is synchronized with the \REF{} command via inserted delays.
Since TRR sampling instances are unknown, we sweep the starting offset of aggressor activations within each round: for a round with $N$ \ACT{}s and aggressor intensity $n$, we vary the offset from 0 to $N - n$.
We also study higher aggressor intensities ($2\times$ to $4\times$ per \TREFI{}) in these multi-\TREFI{} patterns.


\begin{figure}[ht]
\centering
\includegraphics[width=3.3in,height=\paperheight,keepaspectratio]{"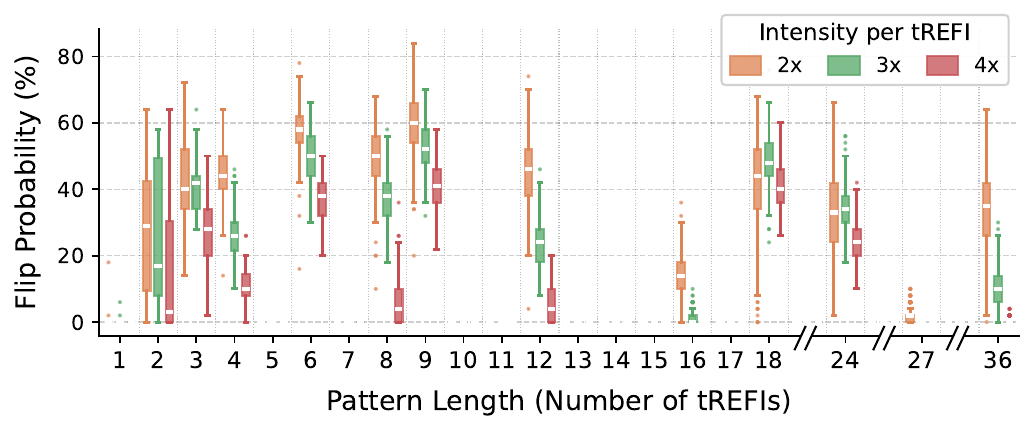"}
\caption{Bit \FlipProbDef{} as pattern length increases to multiple \TREFI{}s. Bit flips are most reproducible (more than 10\% \FlipProbDef{}) at pattern lengths of 3, 4, 6, 9, 18, and 24 \TREFI{}s;
Bit flips stop reliably appearing after 36 \TREFI{}s.}

\label{fig:multitrefi_20trefi}
\end{figure}

\cref{fig:multitrefi_20trefi} shows the \FlipProbDef{} for the same bit flip as we increase the pattern length from 1-36 \TREFI{}s.
The flip is reliably reproduced for pattern lengths of 2, 3, 4, 6, 8, 9, 12, 18, 24, and 36 \TREFI{}s. 
For pattern lengths beyond 36 \TREFI{}s, bit flips are no longer observed reliably (we observed a bit flip just once for a 48 and 72 \TREFI pattern).
Combined with the observation that a majority of the successful pattern lengths divide 72, this suggests that TRR mitigations likely operate every 72 \TREFI{}s in GDDR6.
Notably, higher aggressor intensities ($2$-$4\times$ per \TREFI{}), which fail to reliably produce this bit flip for single-\TREFI{} patterns, consistently reproduce it in multi-\TREFI{} patterns. This shows that non-uniform multi-\TREFI{} patterns can trigger bit flips at higher intensities.



\begin{observation}\refstepcounter{observationcounter}
\textbf{Observation~\theobservationcounter.} Non-uniform multi-\TREFI{} hammering reliably triggers bit flips at \textbf{higher intensities} of at least \textbf{4$\times$ per \TREFI{}}. TRR mitigations likely apply \textbf{every 72 \TREFI{}s}, suggesting that multi-\TREFI{} patterns that divide 72 can reliably trigger bit flips.
\end{observation}

\subsection{Increasing Intensity for \texorpdfstring{$n$}{n} \TREFI{} Patterns}
\label{sec:multref_inc_intensity}
Having identified multi-\TREFI{} patterns that reliably reproduce bit flips (patterns where lengths divide 72), we study how far we can increase aggressor intensity while maintaining reproducibility.

\smallskip
\noindent \textbf{Approach.} 
We start with the more reproducible multi-\TREFI{} patterns that divide 72 \TREFI{} (i.e., 3, 4, 6, 9, 18, and 24 \TREFI{}s long) with a baseline aggressor intensity of $2\times$ \ACT{}s per \TREFI{}, and progressively increase intensity by replacing decoy accesses with aggressor accesses. 
Each step increases the number of aggressor activations, while ensuring no more than one aggressor \ACT{} per warp, to avoid coalescing as discussed in \S\ref{sec:coalescing}.
Since each victim row can have two aggressor rows surrounding it, we first replace decoys greedily with just a repeated single-sided aggressor, generating patterns that increase the intensity from 1 to 7 \ACT{}s per \TREFI{}.
Once all the warps have at least one aggressor, we follow the same procedure to add the \textit{other-side} aggressor to these warps to generate patterns with intensity of 8 to 15 \ACT{}s per \TREFI{}.
For a given pattern, after assigning aggressor activations, all the remaining activations are chosen to target unique decoy rows. For each pattern, we measure its \FlipProbDef{} to identify the higher-intensity patterns 
that reliably induce bit flips.

\begin{figure}[ht]
\centering
\includegraphics[width=3.3in,height=\paperheight,keepaspectratio]{"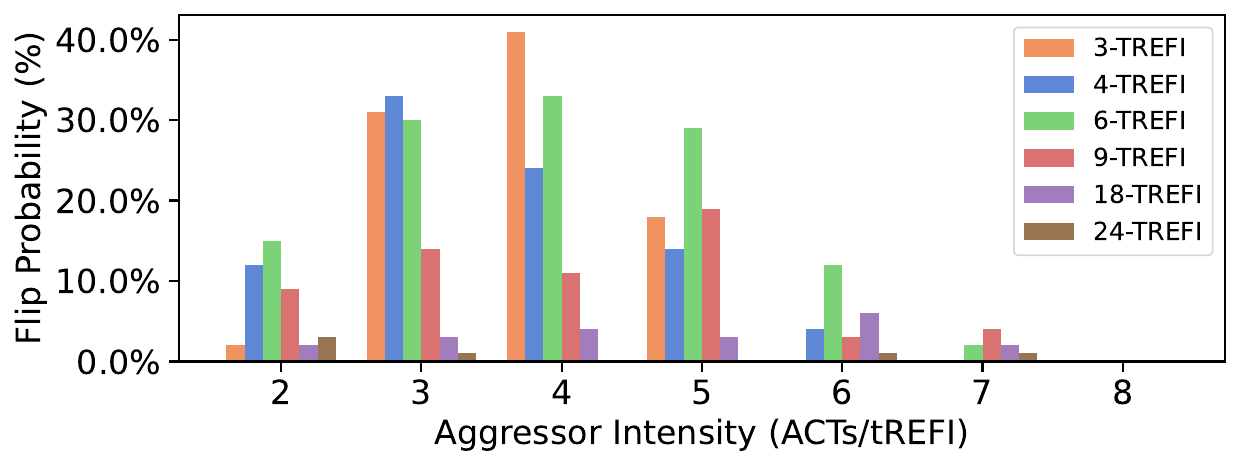"}
\caption{Multi-\TREFI{} hammering with increasing aggressor intensity (\ACT{}s per \TREFI{}) for different pattern lengths (3, 4, 6, 9, 18, 24 \TREFI{} patterns). Our patterns reproducibly trigger bit flips with intensity of up to 7 \ACT{}s per \TREFI{}.}
\label{fig:intensity_analysis}
\end{figure}

\smallskip
\noindent \textbf{Results.}
\cref{fig:intensity_analysis} shows the \FlipProbDef{} across 100 trials for a bit flip as the aggressor intensity increases from 2 to 15 \ACT{}s per \TREFI{} across the pattern, for the reliable multi-\TREFI{} pattern lengths (e.g., lengths that divide 72 \TREFI{} and had $>$10\% \FlipProbDef{} in \cref{fig:multitrefi_20trefi}). 
Shorter pattern lengths, such as 3 \TREFI{} patterns, only trigger bit flips up to an intensity of 5 \ACT{}s per \TREFI{}.
On the other hand, longer pattern lengths such as 24 \TREFI{} patterns trigger bit flips up to an intensity of 7 \ACT{}s per \TREFI{}.
Across all pattern lengths, the \FlipProbDef{} generally declines once \ACT{}/\TREFI{} reaches higher intensities. This is because as the number of aggressor activations increases, the number of instances when decoy rows get sampled that help to evict the aggressor from the TRR sampler decreases. 
Shorter patterns are able to maintain higher \FlipProbDef{} for bit flips at similar intensity, since shorter patterns (e.g., 6 \TREFI{} pattern) have a higher chance of alignment with a TRR mitigation instance (every 72 \TREFI{}s), compared to longer patterns (e.g., 24 \TREFI{}).
We desire both high intensity and \FlipProbDef{} to be able to trigger a larger number of bit flips in Rowhammer campaigns.
Hence, for \papername{}, we choose a 6-\TREFI{} pattern with an intensity of 6.6 \ACT{}s per \TREFI{} as our attack pattern.

\begin{observation}\refstepcounter{observationcounter}
\textbf{Observation~\theobservationcounter.} 
The 6 \TREFI{} pattern triggers flips at high aggressor intensity (6.6 \ACT{}s per \TREFI) reproducibly, making it suitable to trigger a larger number of flips in campaigns.
\end{observation}

\smallskip
\noindent
\textbf{Single-sided vs Double-Sided.}  
We observe that even when using two-sided aggressors in our pattern, the combined intensity of aggressors that triggers flips remains $\leq 7$ \ACT{}s per \TREFI{}. This limit persists regardless of whether we activate only the single-sided aggressor throughout, or equally divide the activations between the two aggressor rows (double-sided).
Thus, the maximum activation intensity appears to be constrained by the GDDR6 TRR implementation. A plausible explanation is that in this TRR implementation, a victim row is definitively refreshed when the combined activations from both adjacent aggressors exceed a threshold: e.g., one-third of total activations within the 72-\TREFI{} mitigation window. Similar TRR mechanisms have been observed in other GPU memories like HBM2~\cite{HBMRowhammer}.
Thus, distributing activations across one or both aggressors does not increase the achievable intensity, indicating that single-sided and double-sided patterns offer similar effectiveness under our high-intensity, non-uniform, multi-\TREFI{} hammering.

\subsection{\papername{}: Putting It Together}
\begin{figure}[ht]
\centering
\vspace{0.1in}
\includegraphics[width=3.3in,height=\paperheight,keepaspectratio]{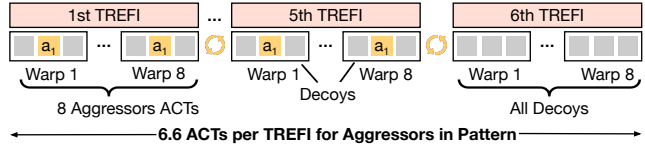}
\caption{GPUThor Hammering Pattern.}
\label{fig:gputhor_pattern}
\end{figure}

\cref{fig:gputhor_pattern} shows the non-uniform, multi-\TREFI{} pattern we use in \papername{} to enable high-intensity Rowhammer attacks on GPUs. The non-uniform hammering uses repeated aggressor activations in distinct warps and uses distinct cachelines within a \TREFI{} to defeat request coalescing. The pattern spans 6 \TREFI{}s, using an aggressor intensity of 6.6 \ACT{}s/\TREFI{} across the entire pattern. The first 5 \TREFI{}s contain repeated aggressor activations (one per warp) and decoy activations, and the last \TREFI{} only contains unique decoy row activations. We observe \papername{} achieves up to 110K aggressor \ACT{}s per \TREFW{}, which is 6.6$\times$ more than prior uniform hammering patterns with GPUHammer~\cite{gpuhammer}.





\begin{table*}[ht]
\centering
\caption{Rowhammer bit flips from campaigns on A4000, A4500, A5000 and A6000, hammering 4 banks each (24 hours / bank). Our campaigns use \papername{} and prior work, GPUHammer~\cite{gpuhammer}.}
\label{tab:rowhammer_result}
\adjustbox{width=\linewidth}{
\begin{tabular}{ccccccccccc}
\toprule
\multicolumn{4}{c}{\textbf{GPUs}} & \multicolumn{1}{c}{\textbf{GPUHammer~\cite{gpuhammer}}} & \multicolumn{6}{c}{\textbf{\papername~ (Our Work)}}  \\
\cmidrule(lr){1-4} \cmidrule(lr){5-5} \cmidrule(lr){6-11}
\textbf{Name} & \textbf{Arch} & \textbf{VRAM} & \textbf{Rows / Bank} & \makecell{\textbf{Total} \\ \textbf{Unique Flips}} & \makecell{\textbf{Total} \\ \textbf{Unique Flips}} & \makecell{\textbf{Avg.} \\ \textbf{Flip / Row}} & \makecell{\textbf{Avg.} \\ \textbf{Flip / Bank}} & \makecell{\textbf{Avg.} \\ \textbf{Flip / GB}} & \makecell{\textbf{Flip / Hour} \\ (Best Data Pattern)} & \makecell{\textbf{Min.} \\ \textbf{HC$_{\text{first}}$}}\\
\midrule

A6000 & GA102 & 48GB GDDR6 & 64K
 & 22 & 57247 (\textbf{2602$\times$})  & 0.22& 14311 & 114K & 1501 &  12.3K\\

\makecell{A5000 } & GA102 & 24GB GDDR6 & 32K & 658
 & 94389 (\textbf{143$\times$}) & 0.76 & 23597 & 377K & 3584 & 15.8K \\

\makecell{A4500 } & GA102 & 20GB GDDR6 & 32K
 & 19 & 18756 (\textbf{987$\times$}) & 0.15 & 4689 & 75K & 882 & 13.6K \\

\makecell{A4000} & GA104 & 16GB GDDR6 & 32K
 & 23 & 18192 (\textbf{790$\times$}) & 0.15 & 4548 & 72K & 838 & 12.4K \\
\bottomrule
\end{tabular}
}
\end{table*}

    


\section{Results}
\label{sec:results}

We evaluate \papername{} on a broad set of NVIDIA GPUs with GDDR6 memories to study its bit-flip behaviors and real-world implications.
Our evaluations focus on answering the following questions:

\begin{enumerate}[leftmargin=*,topsep=0.2em,itemsep=0.15em]
\item Does \papername{} induce significantly more bit flips than prior attacks on a wide range of NVIDIA GPUs? (\S\ref{subsec:vuln_char}) 
\item {Do the increased bit flips with \papername{} reduce the time needed for GPU Rowhammer-based privilege-escalation exploits?} (\S\ref{subsec:time_to_exploit})
\item {Does \papername{} induce multi-bit flips that can potentially defeat ECC on GDDR6 based NVIDIA GPUs?} 
(\S\ref{subsec:multi_bit})
\item { Can \papername{} trigger exploits on GPUs despite NVIDIA-rec\-ommended mitigations like ECC being enabled?} (\S\ref{sec:ecc})
\end{enumerate}

\subsection{Experimental Setup}
\label{subsec:setup}

\noindent\textbf{Target GPUs.} We evaluate \papername{} across four Ampere-class NVIDIA GPUs: the RTX A4000, A4500, A5000, and A6000, spanning the GA102 and GA104 architectures with 16--48\,GB GDDR6 memory. The RTX A6000 is hosted locally, while the A4000, A4500, and A5000 are accessed on the cloud. Our A6000 uses Samsung GDDR6, as confirmed using \texttt{LACT}~\cite{lact}; the cloud GPUs do not expose the low-level permissions required to identify the memory vendor.
All experiments in \S\ref{sec:results} are conducted with ECC {disabled}, reflecting the default in several cloud GPUs (our cloud GPUs do not enable ECC by default, nor provide us the permissions to enable ECC); we enable ECC for the local A6000 for evaluations in \cref{sec:ecc}. We use Ubuntu 22.04 with IOMMU enabled, like prior work~\cite{gpubreach}.

\smallskip
\noindent\textbf{Recovering Address-to-Row Mappings.} 
Like GPUHammer~\cite{gpuhammer}, we reverse-engineer the GPU virtual-address-to-DRAM bank and row mapping in two steps. First, we use timing side-channels on row-buffer conflicts~\cite{gpuhammer,drama} to recover addresses that map to the same bank. Within each bank, we identify addresses mapping to the same row based on row-buffer-hit timing and identify unique addresses mapping to the same bank and consecutive rows. We use this set of addresses, called the \emph{row set}, for hammering campaigns. 

\smallskip
\noindent\textbf{Hammering Configuration.} 
Our hammering kernels use 8 warps $\times$ 3 threads, with each thread hammering one row in a \TREFI{} like GPUHammer~\cite{gpuhammer}; the pattern synchronizes with \REF{} every $n-$\TREFI{} through added delays. We introduce an outer loop for multi-\TREFI{} patterns to hammer a different schedule of activations every \TREFI{} for \papername{}'s non-uniform patterns. Our hammering campaigns sequentially select each row in the bank as aggressor, and hammer using four victim/attacker data patterns: 0xff/00, 0x00/ff, 0xaa/55, 0x55/aa. Our hammering campaigns last for 24 hours/bank on each GPU and we hammer four banks per GPU. 

\smallskip
\noindent\textbf{Exploits.} As a proof-of-concept for exploitation with our bit flips, we use the GPU page-table corruption exploit similar to prior works~\cite{gpubreach, gddrhammer, geforge}. For this, we use the open-sourced GPUBreach exploit code~\cite{gpubreachgithub}. Here, the attacker massages PTEs into target DRAM rows, hammers neighboring aggressors to flip the page-frame-number (PFN) bits of a PTE, and ensures the tampered PTE maps to another PTE via an additional massaging step, to gain arbitrary read/write privileges to the entire memory. For the ECC-enabled settings, we describe our exploit methodology in \S\ref{sec:exploit}.

\begin{table*}[t]
    \centering
    \small
    \caption{Time to mount privilege-escalation exploit using \papername{} compared to GPUHammer~\cite{gpuhammer}.}
    \label{tab:time_to_exploit}
    \begin{tabular}{l|cccc|cccc}
    \toprule
    & \multicolumn{4}{c}{\textbf{GPUHammer~\cite{gpuhammer}}} 
    & \multicolumn{4}{c}{\textbf{\papername{} (our work)}} \\
    \cmidrule(lr){2-5} \cmidrule(lr){6-9}
    \textbf{GPU}
        & \makecell{\textbf{Row Set} \\ 
        \textbf{(min)}}
        & \makecell{\textbf{Bit-Flip Discovery} \\ \textbf{(min)}}
        & \makecell{\textbf{Privilege Escalation} \\ \textbf{(min)}}
        & \makecell{\textbf{Total} \\ \textbf{(min)}}
        & \makecell{\textbf{Row Set} \\ 
        \textbf{(min)}}
        & \makecell{\textbf{Bit-Flip Discovery} \\ \textbf{(min)}}
        & \makecell{\textbf{Privilege Escalation} \\ \textbf{(min)}}
        & \makecell{\textbf{Total} \\ \textbf{(min)}} \\
    \midrule
   A6000 
        & 90 & 1223.6 & 0.3 & 1313.9
        & 0.5 & 0.3 & 0.3 & 1.1 \\
    A5000 
        & 3.1 & 3.6 & 0.2 & 6.9 
        & 0.3 & 0.1 & 0.2 & 0.6  \\ 
    A4500 
        & 8.8 & 237.5 & 0.2 & 246.6 
        & 0.5 & 0.5 & 0.2 & 1.2 \\ 
    A4000 
        & 10 & 403.8  & 0.2 & 414.0 
        & 0.4 & 0.6 & 0.2 & 1.2 \\ 

    \bottomrule
    \end{tabular}
\end{table*}



\subsection{Bit Flip Characterization}
\label{subsec:vuln_char}
\cref{tab:rowhammer_result} shows the results of our hammering campaigns on 4 GPUs (A4000-A6000) targeting 4 DRAM banks each.
While GPUHammer~\cite{gpuhammer} produces 19 to 658 bit flips on these GPUs, \papername{} yields 18,000 to 94,000 unique bit flips.
\papername{} achieves 143$\times$ (A5000) to 2{,}602$\times$ (A6000) more bit flips compared to GPUHammer. 

Across the four GPUs, we observe that the A5000 GPU has the highest vulnerability levels, resulting in 377K bit flips per GB. The A6000 GPU has the second-highest vulnerability level, with 114K bit flips per GB. The A4500 and A4000 have the lowest vulnerability levels with 75K and 72K flips per GB.
The bit flip rates with the GDDR6 DRAM on the A5000 GPU are close to the worst-case vulnerability in DDR4 DRAM with state-of-the-art CPU attacks like Blacksmith (550K flips per GB)~\cite{Blacksmith}, indicating that our non-uniform GPU Rowhammer attacks produce similar worst-case bit flip rates as the best CPU Rowhammer attacks.



On the A5000, the most vulnerable GPU, we observe 3{,}584 flips per hour (one per second) for the best data pattern with \papername{}.
This rate, achieved by hammering a single bank at a time, is almost 180$\times$ faster than the prior best of 20 flips per hour with GDDRHammer~\cite{gddrhammer} and multi-bank hammering (6 banks hammered in parallel).
Across GDDR6 banks of a GPU, the vulnerability levels are similar: e.g., on the A5000, the average (23597 flips per bank) is similar to that of the most vulnerable bank (31797 flips). 
Across the GPUs, we observe an average of 0.15-0.76 flips per row. This makes multiple flips per row and cacheline a common occurrence, and multi-bit flips capable of defeating ECC likely (\S\ref{subsec:multi_bit}).

Across the GPUs, we also measure the minimum activation count to trigger a bit flip (HC$_{\text{first}}$), by re-hammering the bit flips from GPUHammer patterns with activation counts decremented by 100 until the flips stop appearing. All of the GPUs have HC$_{\text{first}}$ values between 12.3K and 15.8K, with no direct correlation between the GPU's overall vulnerability and its HC$_{\text{first}}$ value.



\subsection{Time to Exploit}
\label{subsec:time_to_exploit}
As \papername{} discovers bit flips significantly faster than prior work \cite{gpuhammer}, it directly reduces end-to-end exploit time. 
We demonstrate this using a privilege-escalation exploit that corrupts GPU PTEs via Rowhammer, following prior attacks~\cite{gpubreach,gddrhammer,geforge}.


\smallskip
\noindent \textbf{Setup.} 
We implement a privilege escalation exploit, similar to GPUBreach~\cite{gpubreach}, using their exploit code~\cite{gpubreachgithub}, replacing the default uniform hammering kernel with \papername{}'s non-uniform hammering. 
We evaluate the exploit using both GPUHammer~\cite{gpuhammer} and \papername{} hammering, and measure the total time including the \emph{offline} and \emph{online} phases of the exploit.
The \textit{offline} phase includes the discovery of the \textit{row-set} (addresses mapping to rows of the same bank), and the hammering campaign to find exploitable bit flips; the \textit{online} phase includes massaging the PTE entries to vulnerable rows and hammering the neighbors to escalate privileges~\cite{gpubreach}.

\smallskip
\noindent \textbf{Results.}
\cref{tab:time_to_exploit} shows the total exploit time across four GPUs. 
With GPUHammer, the end-to-end exploit can take up to several hours (21.9 hours on A6000 and 7 hours on A4000). 
The A5000, the most vulnerable GPU, is the exception where the exploit succeeds in 6.9 minutes. 
The main bottleneck in the exploit is the \emph{offline phase}, particularly the exploitable bit-flip discovery, which takes almost 20.4 and 6.7 hours on the A6000 and A4000.
This is because for the exploit to succeed~\cite{gpubreach}, we need the bit flips to be at specific offsets (25 bits of the page frame number in the 8\,B PTE entry). 
This takes only 3.6 minutes on the A5000, the most vulnerable GPU.

In contrast, \papername{} reduces bit-flip discovery time to less than a minute (0.1–0.6 minutes), bringing the total exploit time down to 0.6–1.2 minutes across all the GPUs (10$\times$ to 1200$\times$ lower than GPUHammer). 
This is due to the significantly higher rate of exploitable bit flips discovered by \papername{}. 
This also makes row-set discovery time faster, as \papername{} requires a smaller row-set (tens of rows) given the higher spatial density of its discovered bit flips, compared to GPUHammer that requires larger row sets (tens of thousands of rows across multiple banks). 

The online phase (PTE massaging and final hammering) remains unchanged in GPUHammer and \papername{} (0.2--0.3 minutes) and is a small fraction of the total runtime.

\subsection{Multi-Bit Flips}
\label{subsec:multi_bit}

Beyond increasing the number of bit flips compared to prior works \cite{gpuhammer,gddrhammer,geforge,gpubreach}, \papername{} also provides bit flips with higher spatial density.
\cref{tab:multi_bit_erros} quantifies the number of multi-bit flips observed in \papername{}'s hammering campaigns across GPUs at 8\,B, 16\,B and 32\,B granularity.
These are based on the potential ECC code granularity in GDDR GPUs that store 2\,B ECC per 32\,B data~\cite{IMTSullivan}.

At 16\,B and 32\,B granularities, we observe \textbf{double-bit flips} on all GPUs, with counts increasing at larger granularities. This is most pronounced on the A5000 and A6000, with up to 137 and 26 double-bit flips per bank at 16\,B granularity (276 and 54 within 32\,B), respectively. The A4000 and A4500 exhibit fewer multi-bit flips, consistent with their lower overall vulnerability.
These double-bit flips can potentially result in detectable, uncorrectable errors (DUE) with SECDED ECC, causing data loss or denial of service.

The A5000 also exhibits two \textbf{triple-bit flips} at a 16\,B granularity and four at a 32\,B granularity (two of them also fall within an 8\,B region). These are particularly concerning, as they exceed both the correction and detection capability of SECDED ECC on GDDR-based GPUs, resulting in silent data corruption (SDC). 
Next, we evaluate implications of \papername{}'s bit flips on ECC-enabled GPUs.

\begin{table}[ht]
  \caption{Multi-bit flips across A4000, A4500, A5000, A6000 at 8-, 16- and 32-byte granularity. All the GPUs have \textit{double} bit flips, while the A5000 also has \textit{triple} bit flips.}
  \label{tab:multi_bit_erros}
  \adjustbox{width=3.3in}{
  \begin{tabular}{c|ll|ll|ll|ll|lll}
\toprule
  & \multicolumn{8}{c|}{\textbf{Double Bit Flips}} & \multicolumn{3}{c}{\textbf{Triple Bit Flips}} \\ \cmidrule{2-12}
  & \multicolumn{2}{c|}{A4000} & \multicolumn{2}{c|}{A4500} & \multicolumn{2}{c|}{A5000} & \multicolumn{2}{c|}{A6000} &
  \multicolumn{3}{c}{A5000} \\

 & 16B & 32B & 16B & 32B & 16B & 32B & 16B & 32B & 8B & 16B & 32B \\
\midrule
Bank 1 & 6  & 13  & 5  & 6  & 73  & 138 & 26  & 54 & 1 & 1 & 2  \\
Bank 2 & 2  & 6   & 4  & 11 & 40  & 79  & 15  & 27 & - & - & - \\
Bank 3 & 4  & 10  & -    & -    & 137 & 276 & 7  & 11 & 1 & 1 & 2 \\
Bank 4 & 1  & 3   & 6  & 10 & 56   & 128  & 5   & 10 & - & - & -\\
\bottomrule
\end{tabular}
}
\end{table}

\section{Implications for ECC Protections on GPUs}
\label{sec:ecc}
As \papername{} can induce multi-bit flips beyond ECC's correction capability, we analyze how NVIDIA’s ECC behaves under multi-bit flips by running Rowhammer campaigns using \papername{} with ECC enabled  (\S\ref{subsec:ecc_campaign_setup}), derive insights on NVIDIA's ECC algorithm (\S\ref{subsec:ecc_results}) and discuss potential exploits on ECC-protected GPUs (\S\ref{subsec:ecc_exp}).

\subsection{Reverse Engineering ECC on NVIDIA GPUs}\label{subsec:ecc_campaign_setup}

NVIDIA GPUs use an undisclosed ECC algorithm. Prior work~\cite{IMTSullivan} has speculated that GDDR6 NVIDIA GPUs provision 2B of sideband ECC per 32B data access, resulting in a 6.25\% memory storage overhead. 
However, reverse-engineering the exact ECC implementation on GDDR memories is difficult without tools like logic analyzers and FPGA-based testbeds used in prior CPU DRAM studies~\cite{eccfail,Eccploit}.
Thus, we hammer blindly with ECC enabled and leverage side-channels to infer the ECC scheme details.

\begin{figure}[ht]
\centering
\includegraphics[width=3.3in,height=\paperheight,keepaspectratio]{"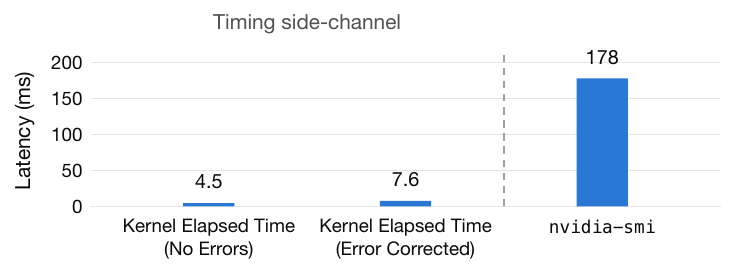"}
\caption{Side channels allowing detection of corrected errors: kernel completion time increases when an error is corrected during execution (4.5\,ms $\rightarrow$ 7.6\,ms); nvidia-smi that also reports corrected error counts has higher latency (178 ms).}
\label{fig:sidechannel}
\end{figure}

\smallskip
\noindent
\textbf{Approach.} 
We enable ECC on our GPUs and execute hammering campaigns with \papername{}. 
In the event of correctable errors, NVIDIA GPUs do not provide logs specifying the bit locations of corrected flips. 
Moreover, we observe that NVIDIA GPUs do not exhibit any timing variations on memory reads when the ECC corrects a bit flip, unlike similar timing side-channels on CPUs observed in prior works~\cite{eccfail,Eccploit}. 
Instead, we discover two \textit{new} side channels: (1) the number of corrected errors reported by \texttt{nvidia-smi} in the ``\texttt{Volatile DRAM Correctable}'' counter, and (2) a timing variation in the completion time of the GPU kernel, due to the delivery of these ECC counter updates to the CPU driver. As shown in \cref{fig:sidechannel}, the kernel completion time shows an increase from 4.5\,ms$\rightarrow$7.6\,ms when an error is corrected during the kernel execution. These kernel execution times are much less than the latency to probe \texttt{nvidia-smi} (178 ms); hence, we use the kernel completion-time side-channel in our campaigns to detect correctable errors.

In the event of detectable but uncorrectable errors (DUE), the CUDA kernel crashes and subsequently requires a GPU reset: fortunately, the GPU driver logs the byte location of such errors to \texttt{syslog}, allowing us to infer their locations. 
In such events, we use a GPU reset to restart our campaigns.
DUEs also trigger a row-remapping on Ampere GPUs~\cite{nvidia_error_management}, which remaps the row where the DUE is detected to a spare row in the bank (up to 8 remaps per bank permitted). 
In our campaigns, we flash the GPU InfoRom with \texttt{nvflash}~\cite{nvflash} before a GPU reset to skip the row-remapping.
We discuss later (\S\ref{sec:exploit}) how an unprivileged attacker can work around row-remapping for exploits with uncorrectable bit flips.


When we detect a correctable error or DUE, to recover the exact \textit{bit} location within a row of the bit flip, we apply \textit{masking} \cite{Eccploit, eccfail, posthammer} similar to prior works. Here, we invert a selected bit in the row in the expected direction of a flip, and try to re-trigger the error by hammering; if it fails to trigger, then we know the bit flip location.

If a triple bit flip is triggered, it surpasses the ECC's detection capabilities and leads to a Silent Data Corruption (SDC); here we observe multiple bits remain flipped even after our side-channel indicates an error was corrected and the kernel does not crash.

\smallskip
\noindent
\textbf{Campaign Setup.}
The A5000 GPU, which is the most vulnerable in our test set, is cloud-based (as are the A4000 and A4500), where the cloud provider does not expose permissions to enable ECC. Thus, we only run campaigns with ECC enabled on the local A6000 GPU.
We run a campaign on 4 banks, where hammering a full bank with ECC enabled to find DUEs takes $\sim$1 day per bank per data pattern. For a campaign that also identifies the bit locations of flips in each DUE and corrected SBE via bit masking, it takes $\sim$1.5 days per bank per data pattern.  

\subsection{Insights on NVIDIA ECC Implementation}\label{subsec:ecc_results} 
In our campaign with ECC enabled on the A6000, we observe 94 DUEs and 1 SDC across four banks. We derive several insights about the NVIDIA ECC based on this. 

\smallskip
\noindent \textbf{ECC Granularity.} Across the campaign, all the DUEs have at least two bits flipped within a 32\,B granularity, indicating that NVIDIA's ECC code is calculated at the granularity of $\leq$32\,B. 
Additionally, we observe some instances of two bit flips within 32\,B data that are both corrected simultaneously, i.e., one bit flip per 16\,B on average that is correctable. 
This suggests that NVIDIA uses a SECDED ECC scheme where 1\,B ECC code protects 16\,B of data. Given that each 32\,B data stores a 2\,B ECC, there are two 16\,B non-contiguous chunks of data, each protected by a 1\,B SECDED ECC.

\begin{observation}\refstepcounter{observationcounter}\label{obs:ecc_behavior}
\textbf{Observation~\theobservationcounter.} NVIDIA ECC protects 32\,B data with a 2$\times$1\,B SECDED code, where each 1\,B ECC provides SECDED protection for 16\,B of data (non-contiguous halves of 32\,B data).
\end{observation}


\noindent
\textbf{Miscorrection on DUEs.}
On a DUE, along with the two bits that flip due to Rowhammer (identified by masking those bits and trying to re-trigger them), we observe additional bits may flip within 32\,B data, indicating the ECC incorrectly attempts correction and mis-corrects the data even on a DUE. 
Across the 94 DUEs observed in the hammered banks, each resulted in 2-13 flips ($\sim$5 on average). In a given DUE, the extra flips due to miscorrection always occur at fixed locations relative to the location of the rowhammer-induced flips, and are independent of the data values. Typically, the original Rowhammer bit flips are preserved in the miscorrected data.

\noindent
\textbf{Delayed DUE Handling by Driver.}
Most importantly, on a DUE, the miscorrected data \textit{remains accessible} from the GPU kernel for up to \textbf{10~ms}. Although no new CUDA API calls can be triggered from the CPU, the corrupted data can be consumed by the CUDA kernel on the GPU and used for page table translations if a GPU PTE is corrupted. We use this insight for our DUE-based privilege escalation exploit in \S\ref{sec:exploit}. The GPU context remains accessible to CUDA kernels for $\sim$10 ms after the DUE, after which all the kernels are terminated, and the GPU is unusable until a full reset.

\begin{observation}\refstepcounter{observationcounter}\label{obs:ecc_race_condition}
\textbf{Observation~\theobservationcounter.} After a DUE, GPU kernels remain \textbf{alive for 10 ms} and can \textbf{consume the corrupted data} or illegally \textbf{access data outside their process} through corrupted PTEs.
\end{observation}

\noindent
\textbf{Triple-Bit Error inducing SDC.}
While a two-bit flip yields a DUE, we also found a
\textit{triple-bit} flip within the ECC granularity that gets mis-corrected without crashing the kernel, resulting in an SDC.
A three-bit error is known to exceed the detection \emph{and} correction capacity of a SECDED-based Hamming code, so its syndrome aliases to that of a single-bit error~\cite{HammingCode}.
We observe NVIDIA ECC exhibits similar behavior, where it flags a triple-bit flip as a \textit{correctable} error, mis-corrects it to end up flipping a fourth bit whose location is fixed deterministically based on the location of the other three flips.
Since the triple-bit flip miscorrections do not crash the kernel, these enable privilege escalation exploits~\cite{gpubreach}, as shown in \S\ref{sec:exploit}.

\begin{observation}\refstepcounter{observationcounter}\label{obs:ecc_triple_flip}
\textbf{Observation~\theobservationcounter.} A \textit{triple-bit} flip within the ECC granularity can induce \textbf{Silent Data Corruption} under NVIDIA's ECC.
\end{observation}

\subsection{Exploits on ECC-Enabled GPUs}
\label{subsec:ecc_exp}

\noindent
\textbf{Denial of Service (DoS) Exploits.}
All our GPUs belong to the GA10x architecture, which does not support Error Containment~\cite{nvidia_error_management}. Thus, a single DUE kills all running processes on the GPU, requiring a full GPU reset to recover, resulting in a DoS. On the A6000 with ECC enabled, we can trigger, on average, 23.5 DUEs per bank within one day of hammering ($\sim$1 DUE / hour). 
Since GPUs in the cloud have been shown to have a Mean Time To Repair (MTTR) of 0.3 hours~\cite{cui2025storygpuscharacterizingresilience} to reset the entire node, this reduces the GPU node's availability under a DoS attack by 22.7\% ($\sim$5.5 hours of downtime per day) and causes data loss due to frequent crashes.

\smallskip
\noindent
\textbf{Abusing Return Merchandise Authorization (RMA) Policy.} NVIDIA GPUs have an RMA policy~\cite{nvidia_error_management} for defective products. 
A GPU becomes RMA eligible when row-remapping fails: each GPU bank has 8 spare rows and each DUE triggers a row remap. If any bank experiences more than 8 DUEs, a row-remapping failure flag is set (visible via \texttt{nvidia-smi}), making the GPU eligible for RMA.
From \cref{tab:multi_bit_erros}, 2 of our 4 GPUs meet this condition, indicating that \papername{} can be used by malicious customers to trigger RMA eligibility and request a free GPU replacement. 
In our ECC-enabled campaign, the A6000 has the row-remapping failure flag set after the 9th DUE, within 18 hours. This poses a risk of warranty abuse by GPU owners, with potential financial implications for GPU vendors.

\smallskip
\noindent
\textbf{Privilege Escalation under ECC Miscorrection.} Observation~\ref{obs:ecc_race_condition} and Observation~\ref{obs:ecc_triple_flip} suggest that ECC miscorrections on DUEs and SDCs can both be exploited for potential privilege escalation attacks.
We discuss this in more detail in \S\ref{sec:exploit}.


\section{Privilege Escalation Exploit with ECC Enabled}\label{sec:exploit}

\subsection{Overview}
\begin{figure}[ht]
\centering
\includegraphics[width=3in,height=\paperheight,keepaspectratio]{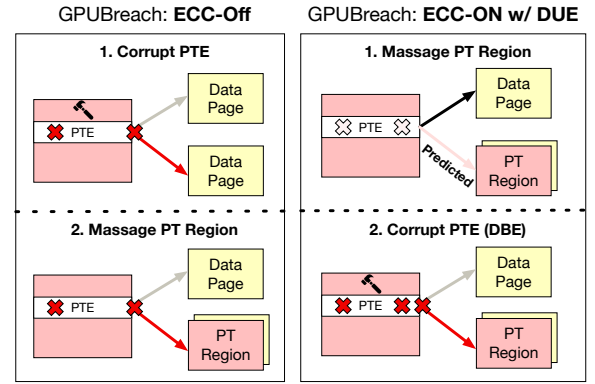}
\caption{Privilege escalation exploit with GPUBreach~\cite{gpubreach} when ECC is enabled using DUEs. We predict the corrupted PTE value and massage a PT into the predicted location in advance, enabling the exploit within 10\,ms of a DUE.}
\label{fig:gpubreach_exploit_difference}
\end{figure}

We demonstrate a privilege-escalation exploit on ECC-enabled GPUs, building on prior exploits~\cite{gpubreach,gddrhammer,geforge} through  Observation~\ref{obs:ecc_race_condition} and Observation~\ref{obs:ecc_triple_flip}.
Prior attacks~\cite{gpubreach,gddrhammer,geforge} follow three key steps: (1) massage page tables (PTs) to vulnerable rows, (2) induce bit flips by hammering neighbors, and (3) massage a new PT at the destination of the corrupted PT entries to control GPU page tables and escalate privileges by accessing and tampering with privileged CPU memory.
However, with ECC enabled, step (3) \textit{can} be infeasible as GPU memory allocations, needed for massaging GPU PTs, are not permitted after a flip in the case of a DUE. 
Thus, for our exploit, we pre-place a PT into a location \textit{predicted} to be the destination of a corrupted PT (\cref{fig:gpubreach_exploit_difference}). After a successful PT  corruption, this allows us to (i) access and modify the pre-placed PTE, altering it to point to CPU memory using the Aperture bits (``A'' in \cref{fig:gputhor_pte}) and (ii) use it to tamper with privileged CPU memory (within 10\,ms before the GPU kernel terminates in the case of a DUE), enabling privilege escalation.
Our threat model assumes a single-tenant setting for the A6000 with ECC enabled. 

\subsection{Offline Profiling for DUEs and SDCs}
\label{subsec:offline-ecc-exploit}

\begin{figure}[ht]
\centering
\includegraphics[width=3in,height=\paperheight,keepaspectratio]{"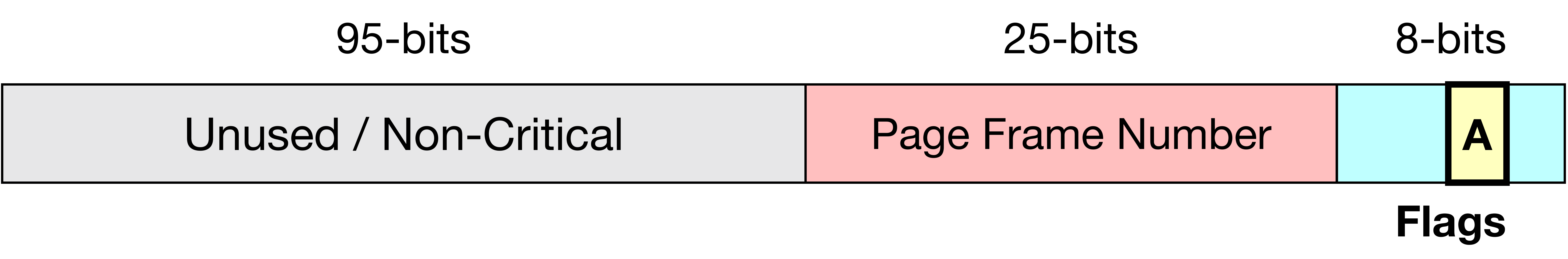"}
\caption{16B PTE Format for 2MB Page Frames. \cite{Pascal_MMU_Format}}
\label{fig:gputhor_pte}
\end{figure}

\smallskip
\noindent
We target 2\,MB PTEs for corruption with bit flips (\cref{fig:gputhor_pte}). 
In these PTEs, a multi-bit flip is exploitable only if (1) some of its bit flips reside in the Page Frame Number (PFN) region of the PTE (25 out of 128 bits) and (2) the tampered PFN ends up in user-accessible memory.
An additional challenge is that miscorrections from either a DUE or an SDC can flip additional bits along with the Rowhammer-induced errors. 
While the original flip locations can be identified easily (\cref{subsec:ecc_campaign_setup}), the miscorrections on DUEs are hard to predict.
Fortunately, out of the 94 DUEs we found in \cref{subsec:ecc_results}, 26 have flips within the PFN 
suitable for our exploit.
The discovered SDC also has exploitable flips within the PFN.

To practically develop an exploit, we need to overcome the remapping of rows on DUEs or induce an exploitable DUE or SDC without triggering the row-remapper. We develop two approaches for this.

\smallskip
\noindent
\textbf{Approach 1: Exhaust Row-Remapper.} After 8 DUEs on a bank, the GPU no longer remaps rows and triggers the remap failure flag on the 9th DUE. Subsequently, any DUEs in the bank persist and become reproducible even after GPU resets and system reboots. Thus, the location of the DUE and the destination of corrupted bit flips in PTEs (i.e., bit flip locations) can be precisely identified. 
We can trigger this state within 18 hours of hammering on our A6000.

\smallskip
\noindent
\textbf{Approach 2: Exhaustive Bit-Search.} Similar to prior ECC-based exploits~\cite{eccfail, Eccploit}, we identify the two or three constituent bit flips of a double or triple bit flip, without triggering them together to avoid a DUE that would trigger a row remap. 
To that end, we use the insight from SoftHammer~\cite{de2025softhammer}, where they show the attacker can gradually increase hammering intensity to avoid collateral damage. Our bit search runs as follows: first, for each row, we gradually increase intensity (2$\times \rightarrow$ 3$\times \rightarrow$ $\dots$) until we observe the first flip in a row. Then, we employ the bit-by-bit search~\cite{eccfail} within each 32B segment of the row, masking each of the 256 bits, to find the second or the third flip without triggering the prior flips.
This process to locate exploitable double and triple bit errors, without triggering DUEs, takes $\sim$4 days on our A6000. 



\subsection{Online Phase: PT Tampering}
Prior work~\cite{gddrhammer,gpuhammer,geforge,gpubreach,tunnels} shows that GPU physical memory allocation is contiguous and reproducible in single-tenant settings. This allows us to deterministically predict the page frame a PTE will reference after tampering, given the flip locations.


\noindent
\textbf{Massaging Page Tables.} 
We leverage \texttt{cuMemMap}~\cite{gddrhammer,gpubreach} to duplicate PTEs with identical physical page frame numbers within a PT region. Given the contiguous virtual-to-physical mappings, the corrupted PTE's destination corresponds to the original virtual address plus the known bit-flip location offset. Prior to hammering, we (1) massage a PT region into the vulnerable row, (2) fill it with duplicated PTEs with \texttt{cuMemMap}, (3) massage a second PT region at the predicted tampered PTE's destination and fill it using \texttt{cuMemMap}.


\noindent
\textbf{Obtaining Arbitrary Read/Write.} 
With the PTEs in place, we hammer the profiled bit flips in the target PTE using \papername{}. 
Then we launch a second GPU kernel that accesses the virtual address obtained from the first \texttt{cuMemMap}, corresponding to the tampered PTE, triggering ECC handling on the flips. If the PTE tampering is successful, we can access the PTEs of the second PT region, and modify them to read/write arbitrary GPU memory or privileged host CPU memory (setting the aperture bits in the PTE~\cite{gddrhammer,gpubreach,geforge}). 




\subsection{Results}
On the ECC-enabled A6000 GPU, we found one SDC and 26 out of 94 DUEs suitable for exploitation (\S\ref{subsec:offline-ecc-exploit}). For the offline phase, we use Approach-2 (\S\ref{subsec:offline-ecc-exploit}), which completes in 4 days. The online phase takes under 2 minutes to obtain an arbitrary read/write primitive, with \texttt{cuMemMap} and TLB flushing accounting for most of that time. 

\smallskip
\noindent
\textbf{SDC-based exploit.} Using the SDC, the privilege escalation to root on the host is straightforward, as the program does not crash after obtaining the arbitrary Read/Write primitive on the GPU memory. By directly leveraging prior work~\cite{gpubreach}, we achieve host-side privilege escalation with ECC enabled on the GPU, even when the IOMMU is enabled.

\smallskip
\noindent
\textbf{DUE-based exploit.}
Using DUEs, once the exploit performs the memory accesses using corrupted PTEs to achieve the arbitrary Read-Write privileges, the GPU kernel terminates within 10\,ms. 
Given this limited time window, this makes directly applying the GPUBreach~\cite{gpubreach} exploit challenging. 
On systems where the IOMMU is disabled by default, a common occurrence in Ubuntu~\cite{markettos2019thunderclap, geforge}, we can still achieve host-side privilege escalation within this brief time window, following prior works~\cite{gddrhammer,geforge}.
Specifically, we modify the pre-placed PTE's aperture bits (``A'' in \cref{fig:gputhor_pte}), to map GPU accesses to CPU memory, and use the arbitrary write primitive to overwrite the process’s credential structure (\texttt{cred}), whose address is assumed to be obtained via side channels~\cite{gpubreach,KASLRFormalDead,liu2023entrybleed}.
By setting \texttt{euid=0}, we get root privileges on the host, similar to prior works~\cite{gddrhammer,geforge}; this persists even after the GPU kernel crashes. 
These steps complete in under 1\,ms, within the available time window, practically achieving host-side privilege escalation even with ECC enabled on the GPU, on host systems with IOMMU disabled.

\section{Discussion and Limitations} 
\textbf{Applicability.} \papername{} induces bit flips on all GA10x GDDR6 GPUs we tested (A4000, A4500, A5000, A6000), but not on the HBM, GDDR6X, or newer-generation GDDR6 GPUs we evaluated (see Appendix~\ref{app:other_gpus}).
These differences may reflect variations in TRR implementations and the use of Refresh Management (RFM) in some GDDR6X-based chips~\cite{not_so_refreshing}.
Nevertheless, our non-coalescing primitives demonstrate the feasibility of non-uniform, multi-\TREFI{} hammering on NVIDIA GPUs, providing a basis for future works to extend such attacks to newer GPUs.

\smallskip
\noindent
\textbf{Newer Reliability Mechanisms in GPUs.} Newer NVIDIA GPUs introduce reliability mechanisms~\cite{nvidia_error_management} with implications for some of our exploits. Server-class Ampere GPUs (A100) and beyond include \textit{Error Containment} and \textit{Dynamic Page Offlining} that
isolate faults to triggering applications; this avoids full GPU resets or crashes of co-tenant processes, improving resilience to denial-of-service attacks. However, these still rely on SECDED-level ECC meaning that SDCs induced by our triple-bit flips still allow privilege escalation attacks to succeed.
\textit{RAS Repair}, introduced in some Blackwell GPUs, replaces a DRAM channel after repeated row-remapping failures within a bank. 
While this makes the attack using DUEs more time-consuming, 
it does not prevent the attack. Finally, newer GPUs using HBM3/e and GDDR7 DRAM support on-die ECC~\cite{JEDEC-HBM3,JEDEC-GDDR7}. While this reduces error visibility, it remains vulnerable to multi-bit flips~\cite{phoenix}.
Future work can explore exploits on these platforms.

\section{Mitigations}
\textbf{Stronger Error Correction.} Deploying stronger ECC
like Chipkill~\cite{Chipkill} can improve resilience to multi-bit flips and enhance Rowhammer protections. However, increasing ECC strength may incur non-negligible storage and bandwidth overheads; existing SECDED ECC in GDDR6 already incurs a 6.25\% memory overhead and up to 10\% slowdown~\cite{IMTSullivan, gpuhammer}. 
This motivates future research on stronger and efficient error correction schemes for GPUs.

\smallskip
\noindent
\textbf{ECC State Monitoring.} 
Just as an attacker can use \texttt{nvidia-smi} reported correctable error counts to identify bit flips (\cref{subsec:ecc_campaign_setup}), an administrator can monitor the ECC-related state to detect Rowhammer activity. A spike in row-remapper activity or number of corrected errors can reveal an attack.
However, such monitoring also has two limitations that may enable evasion:
(1) the row remapper is triggered only by DUEs, not SDCs, and (2) the correctable-error counter becomes unreliable after a few thousand errors and then largely stops incrementing.
We leave a detailed investigation of such monitoring mechanisms for future work.


\smallskip
\noindent
\textbf{Principled Hardware-Level Defenses.} 
Adopting in-DRAM mitigations such as Refresh Management (RFM) and 
Per-Row Activation Counting (PRAC)~\cite{QPRAC,Chronus,MOAT,Panopticon},
or memory-controller-based defenses~\cite{DAPPER,DREAM_defense,AutoRFM, Hydra, AQUA,SRS,RRS}
can be effective mitigations against our attacks.
DRAM integrity protections~\cite{Safeguard,CSIRowhammer, PT-guard} can also detect data corruption via Rowhammer and prevent our exploits. 
\section{Related Work}
\label{sec:related_work}
\textbf{Rowhammer Attacks.} Rowhammer has been studied extensively on CPU DRAM~\cite{RevisitRowhammer,ADeeperLookRowhammerSensitivities_2021,VariableReadDisturb_2025}, with attack patterns bypassing in-DRAM mitigations in DDR3--5~\cite{TRRespass,SMASH,Blacksmith,ZenHammer,phoenix} and on LPDDR~\cite{frigo2018grand,HalfDouble}. ECCploit~\cite{Eccploit} and ECC.fail~\cite{eccfail} further demonstrate that ECC does not inherently prevent Rowhammer on CPU DRAM. These attacks have enabled privilege escalation~\cite{ProjectZeroRowhammer}, cryptographic fault injection~\cite{FFS}, RDMA-based attacks~\cite{ThrowHammer}, browser compromises~\cite{RowhammerJS,SMASH,posthammer}, and attacks on ML models~\cite{prowhammer,PrisonBreak,OneBitFlipAllYouNeed,OneFlipTrojan,bitflipattack,Deephammer,TBD}. Recent attacks have explored other sources of data disturbance such as RowPress~\cite{RowPress} and ColumnDisturb~\cite{ColumnDisturb}. GPUHammer~\cite{gpuhammer} first demonstrated Rowhammer attacks on discrete GPUs, followed by attacks on GPU page tables enabling privilege escalation~\cite{gpubreach,gddrhammer,geforge}. \papername{} brings non-uniform hammering~\cite{Blacksmith} to GPUs, enabling the first GPU Rowhammer attacks that compromise ECC.

\smallskip
\noindent
\textbf{GPU Vulnerabilities.} Prior work has exposed a broad GPU attack surface spanning data leakage, side channels, memory corruption, and privilege escalation. Residual GPU memory can leak renders and ML outputs~\cite{steal_webpage,sorensen2024leftoverlocalslisteningllmresponses}, while microarchitectural side channels and covert channels leak pixels, model parameters, and other sensitive information~\cite{gpuzip,wang2025pixnapping,zhang2025nvbleed,SpyintheGPUBox,LeakyDNN,tunnels,not_so_refreshing}. CUDA memory-safety bugs enable control-flow hijacking and ML model tampering~\cite{GPUMemoryExploitationfunprofit}, and CPU-GPU communication channels can leak data in confidential-computing settings~\cite{GpuCocoDemystified}. GPU Rowhammer~\cite{gpuhammer,gpubreach,gddrhammer,geforge} has expanded this surface to privilege escalation. \papername{} extends this by demonstrating privilege escalation despite ECC protections.

\smallskip
\noindent
\textbf{Rowhammer Mitigations.} Most defenses target CPU DRAM. Software approaches focus on memory isolation~\cite{van2018guardion,brasser2017can,bock2019rip,konoth2018zebram,Siloz,Citadel}, ECC-driven remapping~\cite{di2023copy,nvidia_error_management}, or aggressor rate limiting~\cite{MemoryBandaid_defense, BreakHammer, Blockhammer}; these may extend to GPUs with driver support. Hardware-level defenses, such as tracker-based schemes~\cite{MINT,PRIDE,ProTRR,park2020graphene,Hydra,AutoRFM,mithril,DREAM_defense,ABACuS,CoMeT,Svard_defense_2024,MIRZA_defense,Salt_defense,APT_Rowhammer}, row relocation~\cite{AQUA,Shadow,RRS,SRS}, delayed activations~\cite{Blockhammer}, refresh-generating activations~\cite{REGA_SP23}, page table integrity protection~\cite{PT-guard}, and PRAC-based designs~\cite{QPRAC,MOAT,Chronus,PVAC_defense}, are also applicable to GPU DRAM. However, their latency, bandwidth, and storage overheads on GPUs remain unexplored.

\section{Conclusion}
We present \papername{}, the first high-intensity, non-uniform Rowhammer attack on NVIDIA GPUs. By reverse-engineering coalescing on GPUs and TRR on GDDR6 memories, we design multi-\TREFI{} patterns that achieve 500$\times$–23,500$\times$ more bit flips than prior GPU attacks. 
\papername{} also induces multi-bit flips uncorrectable by the SECDED ECC, and enables denial-of-service attacks and the first privilege-escalation attacks on ECC-enabled GPUs. Our results show that the NVIDIA-recommended mitigation of enabling ECC in GPUs is insufficient
and stronger defenses are needed.

\section*{Acknowledgments}
We thank Prof. Kaveh Razavi for a conversation that inspired this work. 
This research was supported by an NSERC Discovery Grant (RGPIN-2023-04796) and an NSERC-CSE Research Communities Grant (ALLRP-588144-23). Any research, opinions, or positions expressed in this work are solely those of the authors and do not represent the official views of NSERC, the Communications Security Establishment Canada, or the Government of Canada.
%

\bibliographystyle{ACM-Reference-Format}
\bibliography{refs}

\appendix
\section{Open Science} 
Our artifact contains the code for our key results, including the Rowhammer attack campaigns (Table 3), the reverse engineering (Table 2 and Figures 4, 6, 7), and the campaigns with ECC (Section 7).
Our code will be available at \url{https://github.com/sith-lab/gputhor}.


\section{Ethical Considerations} 

\smallskip
\noindent
We responsibly disclosed our findings to NVIDIA and the major cloud providers (Google, Microsoft, AWS) prior to the public disclosure. 
We also notified the cloud GPU provider whose GPUs we found had ECC disabled.
All experiments were conducted on locally owned GPUs or on cloud GPUs within VMs we controlled, without any impact to other users. 
All ECC-enabled experiments (DoS and privilege escalation) were performed only on locally owned GPUs. 


\section{Generative AI Usage}
We used GPT-5.2 and Claude Sonnet 4.6 and Opus 5 for minor grammatical corrections and light stylistic refinements of the manuscript. All such edits were carefully reviewed and verified by the authors.

\section{Evaluation on Other GPUs}
\label{app:other_gpus}
We tested several other GPUs available through cloud providers. For each GPU in \cref{tab:gputhor_generalizability}, we generate a Row Set for 1GB of memory and perform \papername{} across all intensity levels (2$\times$ to 6.6$\times$). However, other than the A4000-A6000 with GDDR6 memory, none have bit flips with our attack patterns, suggesting different TRR implementations that may require more sophisticated hammering or lack of a vulnerability to Rowhammer.

\begin{table}[th]
\centering
\caption{Evaluations on Other GPUs}
\label{tab:gputhor_generalizability}
\adjustbox{width=3.3in}{
\begin{tabular}{l|ccccc|c|c}
\toprule
& \multicolumn{5}{c}{GDDR6} & \multicolumn{1}{|c|}{GDDR6X}  & \multicolumn{1}{c}{HBM2} \\\cmidrule{2-8}
& \makecell{3080 Mobile\\(GA104)} & 
 \makecell{A10\\(GA102)}
&\makecell{A2000-6000 ADA\\(AD102-107)}
& \makecell{L4\\(AD104)}
& \makecell{L40\\(AD102)}
& \makecell{4090\\(AD102)} & \makecell{A30\\(GA100)} \\
\midrule
Flips & No  & No & No  & No & No & No & No \\
\bottomrule
\end{tabular}
}
\end{table}

\section{Data Pattern Dependency}
\begin{figure}[ht]
\centering
\includegraphics[width=3.3in,height=\paperheight,keepaspectratio]{"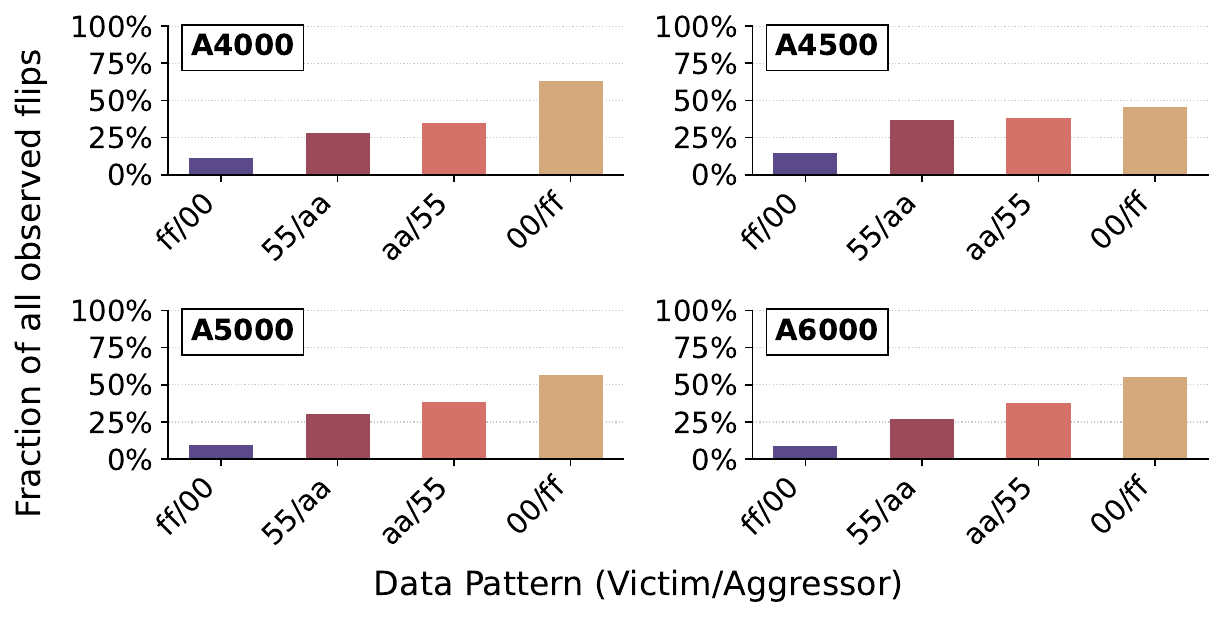"}
\caption{Percentage of Unique Flips Found per Data Pattern for each GPU.}
\label{fig:datapattern_dependency}
\end{figure}

\cref{fig:datapattern_dependency} shows the effect of aggressor and victim data patterns on the number of bit flips in \papername{} campaigns, previously discussed in \cref{tab:rowhammer_result}. Consistent with prior work~\cite{gddrhammer,gpuhammer}, we see that victim/attacker data patterns of 0xff/00 induces the most bit flips, triggering more than half the flips observed in our campaigns.

\begin{figure}[htb]
\centering
\includegraphics[width=3.3in,height=\paperheight,keepaspectratio]{"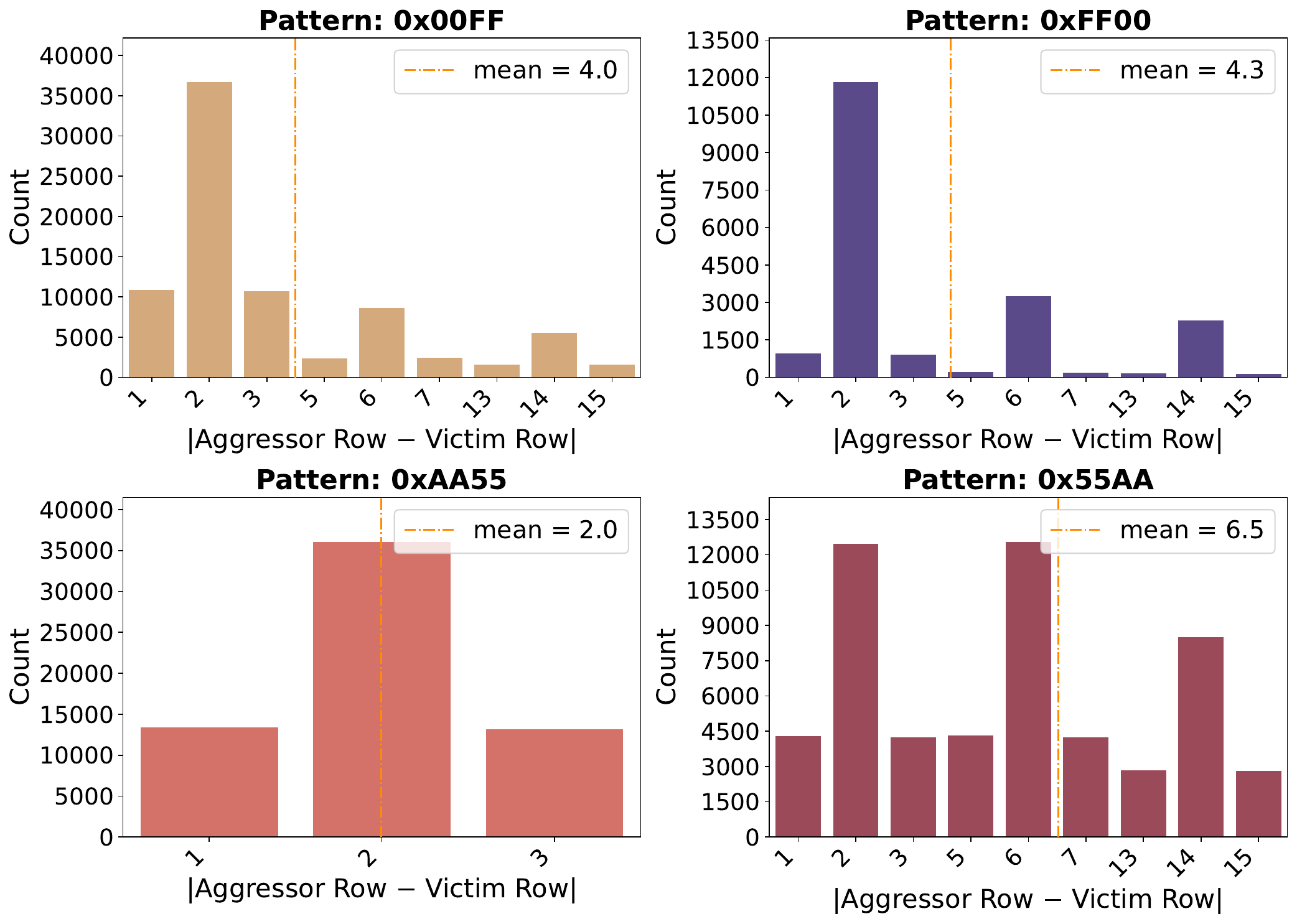"}
\caption{Aggressor-to-Victim Distance distribution for different Data Patterns. Data collected from bit flips occurring across all campaigns and GPUs.}
\label{fig:agg_distance_by_datapattern}
\end{figure}

We revisit the aggressor–victim distance distribution reported in prior work~\cite{gddrhammer,gpuhammer}, measured separately for each data pattern. \cref{fig:agg_distance_by_datapattern} shows that each data pattern yields a distinct distance distribution for bit flips.
Most victim/aggressor data patterns trigger bit flips in victim rows that are logically at most 15 rows away from aggressor rows, validating that there is a non-linear mapping of logical rows to physical rows in the GDDR6 DRAM, as previously observed by GDDRHammer~\cite{gddrhammer}.

However, the victim/aggressor data pattern 0xAA/55 has bit flips only when an aggressor row is within a logical distance of 3. 
This suggests that certain physical rows (where the logical distances between aggressor and victim are larger than 3) are not vulnerable to this data pattern, as they are to other data patterns.
We hypothesize that this could be due to differences in data scramblers used in different parts of the bank.
This data dependence in the Rowhammer susceptibility, linked to spatial locations in the bank, could lead to new side-channels that leak the data in neighboring rows. 
We leave explorations of this for future work.

\newpage

\section{Unique Decoys Sensitivity Analysis}

\begin{figure}[th]
\centering
\includegraphics[width=3.3in,height=\paperheight,keepaspectratio]{"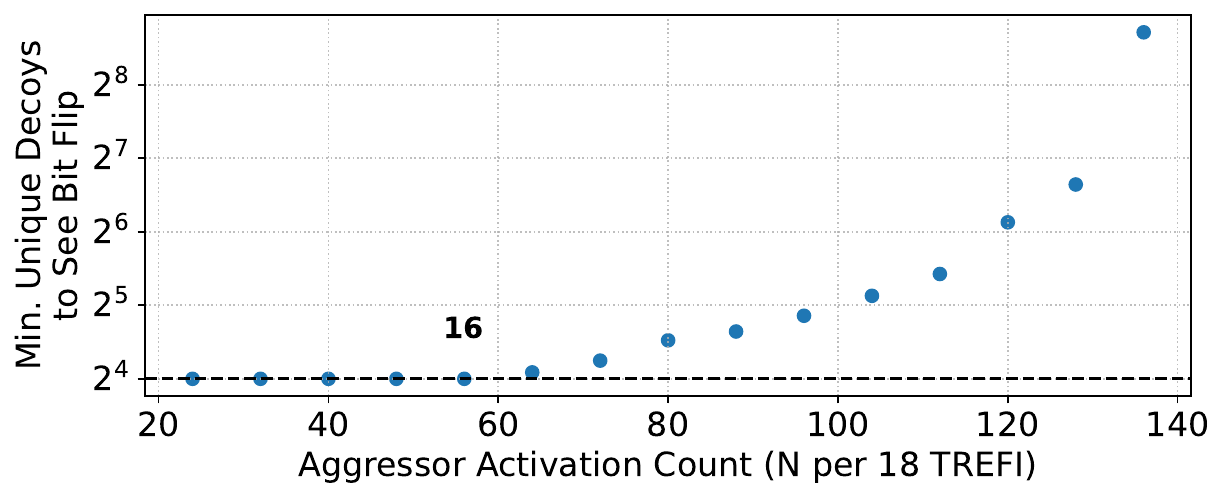"}
\caption{Minimum unique decoy rows required to trigger bit flip for the 18-TREFI pattern when we extend our hammering to higher intensity. At 56 \ACT{}s or less, only 16 unique decoys (the tracker size) are needed, but when the aggressor activation counts are increased to 135 \ACT{}s, up to 400 unique decoy rows are required to trigger bit flips.}
\label{fig:required_dummies_analysis}
\end{figure}

In \cref{sec:multitrefi}, our attack patterns utilize unique, randomly selected decoy rows (one activation per decoy). However, for counter-based TRR designs~\cite{TRRespass}, issuing more than one \ACT{} per decoy row may evict aggressors more reliably from the TRR sampler. To evaluate this, we study whether the frequency of decoy activations influences attack effectiveness. For a given aggressor activation count in a 18 \TREFI{} hammering pattern, we progressively reduce the number of unique decoy rows in the pattern (keeping the total decoy activations constant) until bit flips are no longer observed. Contrary to our hypothesis, \cref{fig:required_dummies_analysis} shows that at higher aggressor activation counts, successful attacks require more unique decoys rather than repeated accesses to the same rows. Moreover, this threshold grows exponentially with increasing hammering intensity. Thus, unique decoys are preferred to fool the TRR in GDDR6 compared to repeated activations to the same decoy.
This suggests that the TRR tracker in Samsung GDDR6 memories is likely counter-less or has a very small saturating counter per entry.

\end{document}